\documentclass[12pt]{article}

\usepackage{url}
\usepackage{authblk}
\usepackage{array}
\usepackage{amsthm,amsmath,bbm,amssymb}
\usepackage[status=draft,inline,index]{fixme}
\usepackage{graphicx,subcaption}
\usepackage[margin=1in]{geometry}
\usepackage[square,numbers, sort]{natbib}
\usepackage{float}
\usepackage{enumitem}
\newtheorem{assumption}{Assumption}
\newcommand{\independent}{\perp\mkern-9.5mu\perp}
\newcommand{\E}{\mathbb{E}}
\renewcommand{\Pr}{\mathbb{P}}

\newcommand{\bZ}{\mathbf{Z}}
\newcommand{\bz}{\mathbf{z}}
\newtheorem{proposition}{Proposition}

\usepackage{tikz}
\usetikzlibrary{arrows.meta,shapes}

\title{Causal inference for N-of-1 trials}
\author[1,2,3]{Marco Piccininni}
\author[3]{Mats J. Stensrud}
\author[4]{\authorcr Zachary Shahn} 
\author[1,2]{Stefan Konigorski}
\affil[1]{Digital Health - Machine Learning Research Group, Hasso Plattner Institute for Digital Engineering, Potsdam, Germany}
\affil[2]{Digital Engineering Faculty, University of Potsdam, Germany}
\affil[3]{Institute of Mathematics, École Polytechnique Fédérale de Lausanne, Switzerland}
\affil[4]{CUNY Graduate School of Public Health and Health Policy, New York, USA}

\date{}

\begin{document}
  \maketitle
\begin{abstract}

The aim of personalized medicine is to tailor treatment decisions to individuals' characteristics. N-of-1 trials are within-person crossover trials that hold the promise of targeting individual-specific effects. While the idea behind N-of-1 trials might seem simple, analyzing and interpreting N-of-1 trials is not straightforward. Here we ground N-of-1 trials in a formal causal inference framework and formalize intuitive claims from the N-of-1 trials literature. We focus on causal inference from a single N-of-1 trial and define a conditional average treatment effect (CATE) that represents a target in this setting, which we call the U-CATE. We discuss assumptions sufficient for identification and estimation of the U-CATE under different causal models where the treatment schedule is assigned at baseline. A simple mean difference is an unbiased, asymptotically normal estimator of the U-CATE in simple settings. We also consider settings where carryover effects, trends over time, time-varying common causes of the outcome, and outcome-outcome effects are present. In these more complex settings, we show that a time-varying g-formula identifies the U-CATE under explicit assumptions. Finally, we analyze data from N-of-1 trials about acne symptoms and show how different assumptions about the data generating process can lead to different analytical strategies.
\end{abstract}

\newpage
\section{Introduction}
The conventional causal estimand in parallel-group randomized controlled trials (RCTs) is the average causal effect in a given population \cite{pearl2009, hernan2019}.
Recent interest in personalized medicine, however, has challenged the reliance on conventional RCTs for identifying clinically relevant causal effects \cite{Schork2018-xs}. Individuals might respond differently to treatment or have different treatment preferences \cite{Schork2018-xs, Lillie2011-gv}. Thus, only knowing that a treatment is effective on average in a given population is insufficient to fully inform a personalized treatment decision \cite{Essentials_ch6, Guyatt1986-je}.
In some cases no RCT is available \cite{Guyatt1986-je, Essentials_ch4}, the patient of interest does not even fulfill the eligibility criteria of the available published RCTs \cite{Guyatt1986-je, Essentials_ch4, Mirza2017-uh,  Essentials_ch11}, or the treatment effect is highly heterogeneous (and therefore the RCT results may not be of direct use) \cite{Guyatt1986-je, Mirza2017-uh, Essentials_ch11, Hogben1953-fe, Schmid2022-oh}.

N-of-1 trials have been proposed as an alternative study design to advance personalized medicine \cite{Schork2018-xs, Lillie2011-gv}. These trials,  which are also called single-case experimental designs \cite{Guyatt1986-je, Manolov2017-fu} or self-controlled experiments \cite{Hogben1953-fe}, aim at assessing the response to an intervention in a specific individual \cite{Schork2018-xs, Essentials_ch6, Guyatt1986-je,  Schmid2022-oh}. The key feature of this design is that the same individual receives the treatment and the comparator in designated, sequential time periods. N-of-1 trials have been defined as “within-patient, randomized, double-blind, cross-over comparisons of a treatment with either placebo or another treatment” \cite{Essentials_ch2}. While there are several definitions of N-of-1 trials, the typical N-of-1 trial is characterized by a multiple crossover design, multiple measurements of the outcome over time, random allocation of the treatment sequence, and blinding \cite{Guyatt1986-je,Schork2018-xs, Schmid2022-oh, Essentials_ch2}. The outcome is frequently chosen to be directly relevant to the specific patient's well-being \cite{Guyatt1986-je, Essentials_ch11}. 

Although N-of-1 trials require careful methodological and ethical considerations and are not universally suitable for all treatments and outcomes \cite{Manolov2017-fu, Schork2018-xs, Mirza2017-uh, Hogben1953-fe, Guyatt1986-je}, they have the potential to allow identification of individual-level rather than population-level causal effects \cite{Lillie2011-gv}. Therefore, N-of-1 trials hold the promise of going beyond clinical decision making informed by effect estimates for the “average patient” \cite{Tabery2011, Hogben1953-fe, Schmid2022-oh, Essentials_ch4, Essentials_ch6, Essentials_ch11}. Results of a single N-of-1 trial may not generalize beyond the individual that is studied \cite{Guyatt1986-je, Mirza2017-uh, Lillie2011-gv}, but series of N-of-1 trials of distinct individuals can be aggregated and analyzed together to yield efficient estimates of population average treatment effects \cite{Schork2018-xs, Mirza2017-uh}. These aggregated N-of-1 trial results have been informally shown to converge towards the results of a multi-crossover RCT \cite{Guyatt1986-je} or a parallel-group RCT \cite{Essentials_ch4, Zucker1997}.
 
The basic idea behind an N-of-1 trial is far from new. Clinicians routinely assign their patients to different therapies sequentially to evaluate individual responses \cite{Mirza2017-uh, Guyatt1986-je, Essentials_ch2, Essentials_ch11, Essentials_ch15}. However, such “trials of therapies” with unstructured pre-post comparison of treatment responses have been shown to be prone to both false positive and false negative results due to placebo effects, desire to please, and natural courses of diseases \cite{Guyatt1986-je, Mirza2017-uh, Essentials_ch2, Essentials_ch11, Essentials_ch15}. In 1986, Guyatt et al.\ \cite{Guyatt1986-je} argued for the N-of-1 randomized trial as a solution to address the validity concerns of uncontrolled individualized trials of therapy and the impossibility of applying RCT results to specific individuals; that is, they proposed N-of-1 trials as a more rigorous version of the common clinical strategy of sequentially trying different treatments in the same individual \cite{Guyatt1986-je, Essentials_ch2, Essentials_ch11}. Similar study designs have been used for decades in experimental psychology to study behavioral interventions \cite{Guyatt1986-je, Manolov2017-fu, Essentials_ch3, Essentials_ch11}, where units are assigned to different interventions over time. The recent emergence of new technologies to automate data collection, such as mobile electronic health devices, wearables, and cell-phone-based diaries, has renewed interest in N-of-1 trials \cite{Lillie2011-gv, Mirza2017-uh} and their integration into the healthcare ecosystem \cite{Selker2021}.

While the idea behind N-of-1 trials might appear to be simple, these trials need to be analyzed and interpreted with care. In this article, we use a formal causal inference framework to clarify some of the methodological subtleties. In particular, N-of-1 trials are often interpreted as one-person RCTs, and confusion frequently arises around the concepts of confounding and the guarantees given by randomization in this study design. 
For example, it has been claimed that N-of-1 trials “do not generally require causal inference methods” if treatment is randomized \cite{daza2019, daza2022}. Some work has characterized randomization as a way to eliminate confounding due to autocorrelation and carryover effects \cite{daza2019, daza2022}, while other work has highlighted the need to account for autocorrelation and carryover effects even in randomized N-of-1 trials \cite{Daza2019n1rt}. Randomization in N-of-1 trials has also been described as a way to balance time-varying confounding factors \cite{Essentials_ch7, Essentials_ch16}, or as a way to minimize confounding and selection bias \cite{Essentials_ch7}. 
Confusion about the foundation of the study design may result in suboptimal analytic strategies and misleading interpretation of the results. 

\subsection{Aim}

The aim of this work is to present an explicit causal framework for N-of-1 trials and formalize intuitive claims from the N-of-1 trials literature. We will define causal estimands for this study design and discuss the assumptions needed for their identification under different data generating mechanisms. This, in turn, motivates estimators.  

We focus on inference from single N-of-1 trials, i.e., when only data from a crossover trial conducted in one individual are available. We will specifically focus on designs where the participant receives treatment according to a treatment schedule that is assigned at the beginning of the study and repeats itself cyclically over time, as these designs are common in practice \cite{Muller2021, Batley2023}. Nevertheless, we will show in Section \ref{sec:treatment_assignment} how our data generating process accommodates other designs for randomized N-of-1 trials in finite time.

The rest of the manuscript is organized as follows: in Section \ref{sec:related_work} we discuss previous research related to our work; in Section \ref{sec:notation_terminology} we introduce our notation and terminology; in Section \ref{sec:basic_scenario} we introduce a  “basic" causal model relying on strong assumptions about the outcome generating process; in Section \ref{sec:causal_estimands} we discuss causal estimands and introduce an individual-specific causal effect that may be of interest in N-of-1 trials; and in Section \ref{sec:inference_basic_scenario} we present identification and estimation results for the basic causal model. We then present a more general and realistic causal model in Section \ref{sec:complex_scenarios}, relaxing some of the strong assumptions of the basic model, and we discuss inference under this relaxed model in Section \ref{sec:inference_relaxed_model}. We discuss the role of randomization in N-of-1 trial inference in Section \ref{sec:randomization}, and analyze data from N-of-1 trials about acne symptoms in Section \ref{sec:data_application}. Finally, we discuss the generalizability of results from a single N-of-1 trial and the aggregation of series of N-of-1 trials to estimate population-level effects in Section \ref{sec:relevance_results} and Section \ref{sec:aggregating_series}, respectively. Concluding remarks are given in Section \ref{sec:conclusion}.

\subsection{Related work}
\label{sec:related_work}
Beyond the classical literature on N-of-1 trials \cite{Essentials}, which does not explicitly consider a causal framework, our work is related to previous research in causal inference. 

In 1999, Robins et al.\ \cite{Robins1999} discussed identification of the subject-specific causal effect in an observational setting with time-varying exposure, confounder, and outcome measurements. They formalized the subject-specific effect as a conditional effect, where the conditioning is  on a potentially unobserved and high dimensional subject-specific baseline variable. Assuming correct model specification and a form of stationarity given the baseline covariate, they observed  that g-computation can be used to consistently estimate subject-specific treatment effects as observation time increases \cite{Robins1999}.

Relying on a similar idea, Daza later proposed a counterfactual methodology specifically for N-of-1 trials \cite{Daza2018-lo}. He discussed causal effect estimation in settings with autocorrelation of outcome measurements, carryover effects, presence of trends and confounding \cite{Daza2018-lo}. He leveraged the assumption of regularity over time (i.e., a type of stationarity assumptions) for the identification of the causal effects \cite{Daza2018-lo, daza2019} and suggested g-computation and inverse probability of treatment weighting for estimation \cite{Daza2018-lo, daza2019}. We consider a different causal estimand from the one proposed by Daza (see Section \ref{sec:randomization}), and rely on different identification arguments to justify our results. 

In 2016, Neto et al.\ \cite{neto2016} developed an instrumental variable approach to estimate causal effects in N-of-1 randomized trials with imperfect compliance. The authors described their assumptions about the data generating process using directed acyclic graphs (DAGs), and relied on some stationarity assumptions. They proved identifiability of the causal effect in a sequential randomized experiment under additivity assumptions, assuming that the causal effect was constant over time. More recently, Qu et al.\ \cite{qu2023} developed an alternative two-stage Bayesian latent structural model approach for the same purpose, using study treatment randomization as an instrumental variable and assuming randomization only at the beginning of the treatment periods. In our work, we do not require strong parametric assumptions for identifiability, and rely on a frequentist framework for inference. However, we do not specifically address issues with non-compliance, which could be an extension in future work. 

Our work is also related to the broader field of causal inference for time series data, see, e.g., Runge et al.\ \cite{Runge2023} and Eichler and Didelez \cite{Eichler2009}. Particularly important is the work from Bojinov and Shephard \cite{Bojinov2019}, who, relying on randomization-based inference, defined estimators for a class of causal estimands in the setting of time series experiments in which the treatment is randomized at each time point. More recently, also based on randomization inference, Malenica et al.\ discussed anytime-valid inference in N-of-1 trials \cite{Malenica2023}, while Liang and Recht \cite{Liang2023} discussed identification and estimation of causal effects in N-of-1 trials. Liang and Recht assumed a linear time-invariant outcome model to identify causal effects when a very long carryover of the treatment is expected. Unlike Liang and Recht \cite{Liang2023}, we do not invoke specific functional relationships for identifiability and do not present identification results for scenarios where the full treatment history affects directly the outcome at any given point.
Furthermore, unlike Bojinov and Shephard \cite{Bojinov2019}, Malenica et al.\ \cite{Malenica2023} and Liang and Recht \cite{Liang2023}, we do not rely on randomization-based inference but on a superpopulation framework. We do not require randomization at all times; we show that our results also hold in the absence of randomization.

\section{Notation and terminology}
\label{sec:notation_terminology}

We will use the term  “N-of-1 trial" to refer to a single N-of-1 trial \cite{Essentials_ch12, Yang2021} and “series of N-of-1 trials" to refer to a situation where data from multiple N-of-1 trials is available to the investigator \cite{Yang2021, Senn2024}. In this work we focus on inference from data of a single individual (i.e., an N-of-1 trial). Broadly, we consider individuals as i.i.d.\ draws from the same near-infinite superpopulation. Therefore we will describe the data generating process considering a sample of $i=1,...,n$ individuals with data collected at $k=1,...,t$ time points, with $n \geq 1, t \geq 2$.  We will use capital letters to indicate random variables (e.g., $X$) and calligraphic letters to indicate their domain ($\mathcal{X}$). Lowercase letters will indicate random variable instantiations or constants. Subscripts will be used to indicate measurements for an individual $i$ at time $k$ ($X_{i,k}$), but, to avoid clutter we will omit the individual subscript from the random variables when the interpretation is unambiguous. Superscripts will indicate counterfactuals (see Section \ref{sec:counterfactual_outcomes}). We will use overlines to indicate the history of all measurements for the same variable until time $k$ (i.e., $\bar{X}_k$), and use $\bar{X}$ to indicate the full history $\bar{X}_t$. Unless explicitly stated, the notation $\bar{x}_b$ will indicate a vector of size $b$ with all elements equal to $x$, while the letter $v$ will be used to indicate generic constant vectors of size $b$ (i.e., $\bar{v}_b$).

In this article, we primarily consider a specific N-of-1 trial design where the participant receives treatment according to a treatment schedule that is assigned at the beginning of the study and repeats itself cyclically over time. We use the term “treatment schedule" to indicate the treatment sequence that repeats itself over time, and the term “cycle" to indicate the span of time covered by one treatment schedule. The usage of these terms may differ in other works on N-of-1 trials and guidelines \cite{Porcino2020}.

\section{Basic scenario} \label{sec:basic_scenario}

\subsection{Treatment assignment mechanism}
\label{sec:treatment_assignment}

Let $q\geq 2$ represent the length of a cycle defined by the investigator and $t \geq q$ be the length of the study. We always consider $q$ as fixed and predetermined. Let $\textbf{Z} \in \mathcal{Z}$ be a random vector indicating the treatment schedule for an individual, containing $q$ binary elements $Z_1,...Z_q$.  Let $A_{k}$ be a binary variable that indicates whether the individual receives treatment ($A_{k}=1$) or control ($A_{k}=0$) at time $k$. We will assume that the treatment status is determined by the state of $\textbf{Z}$ according to the deterministic function
\begin{equation} \label{eq:A1}
\begin{split}
    A_k = f^{A}(\textbf{Z},k)
    \hspace{7pt} \text{ with} \hspace{7pt}  f^{A}(\textbf{Z},k)=Z_{((k-1) \text{ mod } q) + 1}.
    \end{split}
\end{equation}
That is, $A_k$ is assigned to be the element of $\textbf{Z}$ in position $((k-1) \text{ mod } q) + 1$, with mod indicating the modulo operation. We consider $\textbf{Z}$ to be an exogenous random variable (in Section \ref{sec:randomization} we also discuss a setting where $\textbf{Z}$ is not exogenous). We are, however, agnostic about the number of possible treatment schedules, the sequence of the treatments, and the probability distribution of $\textbf{Z}$. We assume that participants fully adhere to the assigned treatment schedule until the end of the study and only consider possible treatment schedules where both treatment and non-treatment are assigned at least once
\begin{equation} \label{eq:Z1}
\mathcal{Z} \subseteq \bigg\{ \textbf{z}=(z_1,...,z_q) \bigg\rvert 0<\sum^q_{k=1}z_{k} < q \bigg\}.
\end{equation}
To fix ideas, consider an investigator who chooses to have $q=3$ time points in a cycle and to randomize the participant to one of two possible treatment schedules from $\mathcal{Z} = \{(1,0,1),(0,1,0)\}$. The investigator designs a study with $t=6$ time points, which could be a six day long trial with measurements once daily. Suppose that a study participant was randomized to the treatment schedule $\textbf{z}=(1,0,1)$, then their treatment status during the study, determined by Equation \eqref{eq:A1}, is $\bar{a}=(1,0,1,1,0,1)$. Thus, the sequence of assigned treatments is defined by iteratively repeating the elements of $\textbf{z}$ until $t$ values are produced. According to condition \eqref{eq:Z1}, schedules $(1,1,1)$ and $(0,0,0)$ cannot be considered by the investigator. Equation \eqref{eq:A1} and condition \eqref{eq:Z1} might appear unnecessarily complex, but they describe a treatment assignment strategy that is flexible enough to accommodate a wide range of designs for randomized N-of-1 trials in finite time. In particular, this treatment assignment strategy can represent, in finite time, any N-of-1 trial design in which the treatment sequence is exogenous and assigned at baseline. Indeed, if $t$ is finite, setting $q=t$ allows us to represent any treatment sequence of size $t$, except sequences that assign an individual to be always treated or always non-treated. In Table \ref{tab:1}, we give examples of how commonly used N-of-1 designs can be represented using our notation. The same design can be represented in different ways depending on the chosen value for $q$, and we only report configurations for the smallest possible value of $q$.

\begin{table*}
\caption{Representation of examples of commonly employed N-of-1 designs}
\label{tab:1}
\begin{tabular}{ | m{6.5cm} | m{2.2cm}| m{1cm} | m{3cm} | m{1cm} |} 
  \hline
  Description & Design name in the literature & Cycle length ($q$) & Treatment schedule ($\textbf{Z}$) & Study length ($t$) \\ 
  \hline
  \hline
  7 days of no-treatment followed by 7 days of treatment & AB & 14 & $(\bar{0}_7 ,\bar{1}_7)$ & 14 \\ 
  \hline
  Alternate 7 days of no-treatment with 7 days of treatment for 4 weeks & ABAB & 14 & $(\bar{0}_7 ,\bar{1}_7)$ & 28 \\ 
  \hline
  10 days of no-treatment, 10 days of treatment, 10 days of no-treatment & ABA & 20 & $(\bar{0}_{10} ,\bar{1}_{10})$ & 30 \\ 
  \hline
  7 days of no-treatment, 14 days of treatment, 7 days of no-treatment & ABBA & 21 & $(\bar{0}_{7} ,\bar{1}_{14})$ & 28 \\ 
  \hline
  After 1 day of treatment and 1 day of no-treatment, randomly choose the treatment status for every day & - & 14 & $(1,0,\eta)$ with $\eta$ uniformly drawn from $\{0,1\}^{12}$ & 14 \\ 
  \hline
\end{tabular}
\end{table*}

\subsection{Data generating mechanism for the outcome}
\label{sec:outcome_mechanism}

We assume that the outcome $Y_{k}$ is measured at each $k$, and that $Y_{k}$ is generated according to the following structural equation
\begin{equation} \label{eq:Y1}
\begin{split}
Y_{k} = f^Y(A_{k}, U, \varepsilon_{k}),
\end{split}
\end{equation}
with $f^Y$ being a function that does not change across time. Further, $U$ represents an unmeasured exogenous random variable with distribution $F_U$, and $\bar{\varepsilon}=(\varepsilon_1,...,\varepsilon_t)$ is a vector of exogenous unmeasured i.i.d.\ variables drawn from a distribution $F_{\varepsilon}$ that does not vary over time. To simplify the presentation, we describe the setting where $Y_k$ is discrete, but our reasoning can be easily adapted to the continuous case. Thus, we are assuming that the outcome variable measured at time $k$ is fully determined by three quantities: 1) the treatment received by the individual at time $k$, 2) the state of the variable $U$ for the individual, and 3) the instance of the noise variable $\varepsilon_{k}$ for the individual. In particular, we are excluding that $\mathbf{Z}$ has a direct effect on $Y_k$ and that the states of $A_k$, $Y_k$ and $\varepsilon_{k}$ have an effect on the outcome at a later point in time. We will elaborate more on the interpretation and assumptions of this causal model in the next section.

\subsection{Interpretation of the basic causal model}
\label{sec:interpretation_causal_model}

In Sections \ref{sec:treatment_assignment} and \ref{sec:outcome_mechanism}, we relied on structural causal models to represent our data generating process. The structural causal model framework was inspired by Laplace's quasi-deterministic conception of causality \cite{pearl2009}. Laplace considered the world to be deterministic, attributing observed “chance” to the investigator's ignorance \cite{Laplace2012-ov}. In this work, we commit to this interpretation and consider randomness as induced by the fact that certain variables are unobserved \cite{pearl2009}. This means that the outcome value at any time $Y_{k}$ can be exactly derived if $f^Y$, $f^{A}$, and the determination of the exogenous variables $\textbf{Z}$, $U$, and $\bar{\varepsilon}$ are known. While the meaning of the treatment schedule $\textbf{Z}$ is clear, interpreting the other exogenous variables requires some care.

Consider the outcome variable $Y_k$ to be a health-related characteristic of an individual (e.g., systolic blood pressure). The structural equation in \eqref{eq:Y1} implies that there are only three possible sources determining the outcome state at time $k$: the treatment $A_k$ assigned at time $k$, an exogenous variable $\varepsilon_k$ that affects \textit{only} the outcome at time $k$, and an exogenous variable $U$ that affects \textit{every} measurement of the outcome during the study. It may be convenient to think of $U$ as the set of characteristics of the individual at the beginning of the trial that affect the outcome at all time points. While some variables affecting the outcome at all time points are usually measured at baseline (e.g., sex), the full set of variables comprising $U$ is typically unknown or cannot be measured. Indeed, depending on what the outcome of interest is, $U$ may acquire only a few values (e.g., the presence/absence of a genetic mutation), or have very large support and summarize every aspect of the individual's existence (e.g., the state of every single cell in the body). In this sense, $U$ can be informally interpreted as an individual-specific variable, and can represent the source of effect heterogeneity of the treatment. 
The variable $U$ is analogous to the $\sigma$ variable introduced by Robins et al.\ \cite{Robins1999} to define the conditioning set of their individual-specific causal effects. Indeed, when we conduct an N-of-1 trial, we consider only one individual $i$ and therefore we are implicitly conditioning on a specific value of $U=u_i$. The variable $U$ will be different depending on the specific study setting, and its composition will have consequences on the plausibility of statistical assumptions and on the transportability of the N-of-1 trial results, as we further consider in Sections \ref{sec:inference_basic_scenario} and \ref{sec:relevance_results}.

According to Equation \eqref{eq:Y1}, $U$ and $A_k$ are not necessarily sufficient to perfectly predict the outcome. We are, indeed, allowing for the existence of exogenous variables $\bar{\varepsilon}$ such that every $\varepsilon_k$ only affects the outcome at time $k$. Suppose we let each $\varepsilon_k$ denote the external factors at time $k$ that affect the outcome measurement at that time. Consider an N-of-1 trial that includes an individual with hypertension, where the outcome of interest is the effect of two different types of anti-hypertensive treatments on systolic blood pressure. Because the study only includes a single individual we can interpret all observations as conditioned on $U=u_i$, that is, conditioned on all common causes of the outcome measurements during the study. For example, $u_i$ may represent information about demographics, systolic blood pressure before the start of the trial, detailed information on diet before the trial, all diseases the participant already had, all medications the participant took, etc. However, despite knowledge of $U$ and the treatment status at time $k$, the systolic blood pressure measurement at time $k$ is not fully determined, as other variables affecting this specific measurement may also exist. For example, at time $k$, a car accident might happen nearby the individual or a loud thunderstorm may begin. These are external factors ($\varepsilon_{k}$) that affect the systolic blood pressure measurement at that time ($Y_{k}$). Here we assume all the random variables $\varepsilon_{k}$ for $k \in \{1,...,t\}$ to be independently drawn from the same distribution.

Because $f^Y$ is defined to be constant in time, we are assuming that the outcome is generated in the same way over time. This also implies that, according to \eqref{eq:Y1}, there are no unmeasured common causes of a proper subset of the outcome measurements. That is, either an unmeasured cause of $Y_k$ affects only the measurement at time $k$ or it affects all $\bar{Y}$ measurements. However, we are agnostic about the specific functional relationship $f^Y$. It is therefore possible to imagine a function $f^Y$ such that the outcome is generated in completely different ways depending on the value of $\varepsilon_k$. For example, $U$ may alter the value of the outcome measurement $Y_k$ only if $\varepsilon_k$ has a specific value, or $Y_k$ could be a linear function of $A_k$ and $U$ when $\varepsilon_k=1$ and a non-linear function when $\varepsilon_k=0$. 

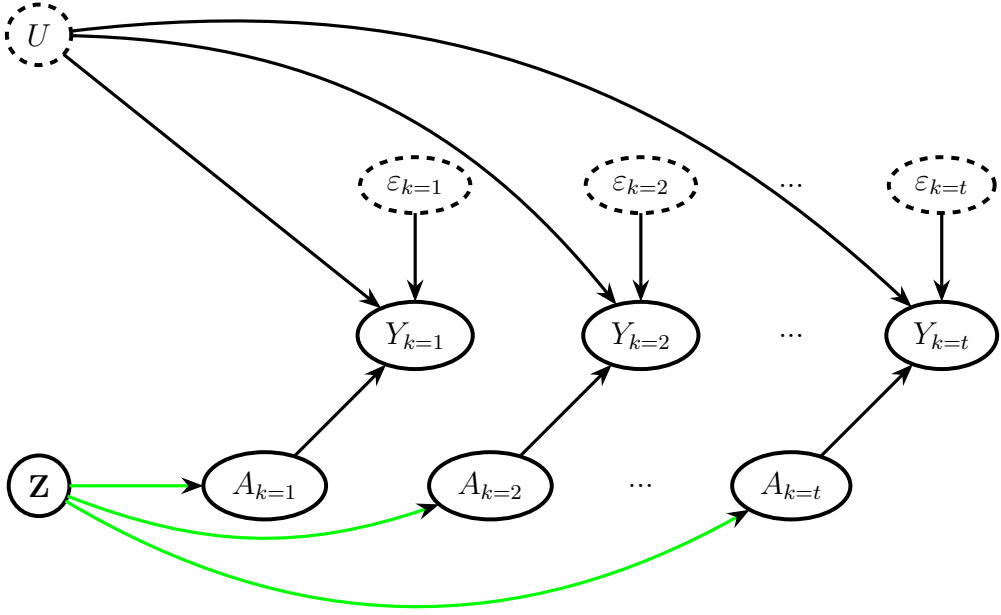
\begin{figure}
    \begin{minipage}{1\textwidth}
        \centering
                \begin{tikzpicture}
                    \tikzset{line width=1.5pt, outer sep=0pt,
                    ell/.style={draw,fill=white, inner sep=4pt,
                    line width=1.5pt}};

                    \node[name=Z,ell,  shape=ellipse] at (0,0) {$\textbf{Z}$};
                    \node[name=U,ell,  shape=ellipse, dashed] at (0,6) {$U$};
                    \node[name=A_1,ell,  shape=ellipse] at (3,0) {$A_{1}$};
                    \node[name=Y_1,ell,  shape=ellipse] at (5,2) {$Y_{1}$};
                    \node[name=epsilon_1,ell,  shape=ellipse, dashed] at (5,4) {$\varepsilon_{1}$};
                    \node[name=A_2,ell,  shape=ellipse] at (6,0) {$A_{2}$};
                    \node[name=Y_2,ell,  shape=ellipse] at (8,2) {$Y_{2}$};
                    \node[name=epsilon_2,ell,  shape=ellipse, dashed] at (8,4) {$\varepsilon_{2}$};
                    \node[name=spa, draw=none] at (8,0) {...};
                    \node[name=spa2, draw=none] at (10,2) {...};
                    \node[name=spa3, draw=none] at (10,4) {...};
                    \node[name=A_t,ell,  shape=ellipse] at (10,0) {$A_{t}$};
                    \node[name=Y_t,ell,  shape=ellipse] at (12,2) {$Y_{t}$};
                    \node[name=epsilon_t,ell,  shape=ellipse, dashed] at (12,4) {$\varepsilon_{t}$};

                     \begin{scope}[>={Stealth[black]},
                                  every edge/.style={draw=black,very thick}]
                        \path [->] (Z) edge[color=green] (A_1); 
                        \path [->] (Z) edge[color=green, bend right=20] (A_2); 
                        \path [->] (Z) edge[color=green, bend right=30] (A_t); 
                        \path [->] (A_1) edge (Y_1);
                        \path [->] (A_2) edge (Y_2);
                        \path [->] (A_t) edge (Y_t);
                        \path [->] (epsilon_1) edge (Y_1);
                        \path [->] (epsilon_2) edge (Y_2);
                        \path [->] (epsilon_t) edge (Y_t);
                        \path [->] (U) edge (Y_1);
                        \path [->] (U) edge[bend left=25] (Y_2);
                        \path [->] (U) edge[bend left=25] (Y_t); 
                    \end{scope}
                \end{tikzpicture}
\end{minipage}
    \caption{Directed Acyclic Graph representing the causal relationships between the assigned treatment schedule ($\textbf{Z}$), the received exposure ($A_k$), the outcome ($Y_k$), the individual-specific random variable ($U$) and the time-varying noise variables ($\varepsilon_k$) in the basic scenario for $t>2$. Nodes represent variables, directed edges represent causal relationships. Dashed nodes represent unobserved variables and green edges represent deterministic causal relationships.}
    \label{fig:dgm1}
\end{figure}

The assumed causal model described in the previous sections is illustrated by a DAG \cite{hernan2019, pearl2009} in Figure \ref{fig:dgm1}. However, the assumption of a fixed $f^Y$ and $F_\varepsilon$ over time, as well as the chosen functional relationship for $f^{A}$, are additional non-graphical assumptions, not represented in the DAG.

\subsection{Counterfactual outcomes and stationarity} \label{sec:counterfactual_outcomes}

Define the counterfactual outcome $Y^{\bar{a}_k=\bar{v}_k}_{k}$, as the outcome we would have observed at time $k$ if the exposure history until time $k$ was externally forced to be equal to $\bar{v}_k \in \{0,1\}^k$.
We further invoke a classical causal consistency assumption \cite{hernan2019},
\begin{equation} \label{eq:consistency}
  Y_{k} = Y^{\bar{a}_{k}=\bar{v}_{k}}_{k} \text{ when } \bar{A}_{k}=\bar{v}_{k}
\end{equation}
for all $k$. The assumption in \eqref{eq:consistency} requires that the counterfactual outcome for the intervention forcing the treatment history $\bar{v}_{k}$ is equal to the observed outcome when $\bar{v}_{k}$ is the observed treatment history.
The structural equation in \eqref{eq:Y1} implies that the counterfactual outcome at time $k$ is affected by treatment interventions only through the intervention at time $k$. That is, for every $k$ and $\bar{v}_k \in \{0,1\}^{k}$,
\begin{equation} \label{eq:carryov1}
Y^{\bar{a}_k=\bar{v}_k}_{k} = Y^{a_k=v_k}_{k},
\end{equation}
with $v_k$ being the last element of $\bar{v}_k$ and $Y^{a_k}_{k}$ indicating the counterfactual outcome at time $k$ when intervening only at time $k$. The equality in \eqref{eq:carryov1} follows immediately from $f^Y(v_{k}, U^{\bar{a}_k=\bar{v}_k}, \varepsilon^{\bar{a}_k=\bar{v}_k}_{k})= f^Y(v_{k}, U, \varepsilon_{k})$, with $U$ and $\varepsilon_{k}$ being the naturally occurring values. Equation \eqref{eq:carryov1} states that there is no carryover effect of the treatment: treatment status at time $k$ does not have any effect on the outcome at times after $k$. Moreover, because we are assuming that $f^Y$ does not change across time and the individual-time specific noise variables are i.i.d,
\begin{equation} \label{eq:stationarity1}
\mathbb{P}(Y^{a_1=x}_{1}=y \mid U=u) = \mathbb{P}(Y^{a_k=x}_{k}=y \mid U=u) 
\end{equation}
for every $k \in \{1,...,t\}$, $x \in \{0,1\}$, $y \in \mathcal{Y}$ and $u \in \mathcal{U}$. Thus, we say that the distribution of the counterfactual outcome is \textit{strongly stationary} across time conditionally on $U$.
The strong stationarity assumption in \eqref{eq:stationarity1} also implies that the expected conditional counterfactual outcome is constant over time, that is,
\begin{equation} \label{eq:stationarity1_expectation}
\mathbb{E}(Y^{a_1=x}_{1} \mid U=u) = \mathbb{E}(Y^{a_k=x}_{k} \mid U=u) 
\end{equation}
holds for every $k \in \{1,...,t\}$, $x \in \{0,1\}$, and $u \in \mathcal{U}$. Similarly, under strong stationarity \eqref{eq:stationarity1}, the conditional variance of the counterfactual outcome is constant over time, and we can define $\sigma^2(x,u)=Var(Y^{a_k=x}_{k} \mid U=u)$ for $k \in \{1,...,t\}$.

Define $k^{\textbf{z}}_x \in \{k_* \in \{1,...,t\} \mid f^{A}(\textbf{z},k_*)=x\}$ to 
be a generic time point in which treatment $x$ is administered under treatment schedule $\textbf{Z}=\textbf{z}$. Because
\begin{equation*} 
\begin{split}
&\mathbb{P}(Y_{k^{\textbf{z}}_x}=y \mid A_{k^{\textbf{z}}_x}=x, U=u, \textbf{Z}=\textbf{z})\overset{\eqref{eq:consistency}}{=} \mathbb{P}(Y^{a_{k^{\textbf{z}}_x}=x}_{k^{\textbf{z}}_x}=y \mid A_{k^{\textbf{z}}_x}=x, U=u, \textbf{Z}=\textbf{z}) \\ &\overset{Y^{a_k}_{k} \independent A_k, \textbf{Z} \mid U}{=} \mathbb{P}(Y^{a_{k^{\textbf{z}}_x}=x}_{k^{\textbf{z}}_x}=y \mid U=u),
\end{split}
\end{equation*}
it follows from Equation \eqref{eq:stationarity1} that the distribution of the outcome at all times under a particular value of treatment is constant when conditioning on $U=u, \textbf{Z}=\textbf{z}$. Therefore, under consistency \eqref{eq:consistency} and the independencies encoded in the DAG in Figure \ref{fig:dgm1}, the consequences of the strong stationarity assumption \eqref{eq:stationarity1} are falsifiable by examining whether the distribution of the outcome is stable over time for an individual, considering only times in which the same treatment is received. A similar argument can be made for condition \eqref{eq:stationarity1_expectation}. As a corollary, we also have that $\sigma^2(x,u)=Var(Y_{k^{\textbf{z}}_x} \mid A_{k^{\textbf{z}}_x}=x, U=u, \textbf{Z}=\textbf{z})$, and therefore this variance can be estimated from observed data if sufficient measurements are available (see discussion in Section \ref{sec:inference_basic_scenario}).

Suppose that we are interested in testing whether the basic causal model is compatible with the observed data. The treatment assignment mechanism described in Section \ref{sec:treatment_assignment} is immediately verifiable since it is directly controlled by the experimenter. To assess whether the outcome generating process is compatible with the one presented in Section \ref{sec:outcome_mechanism}, we could test whether the outcome distribution under the same level of treatment changes over time. For example, consider all the time points $k^{\textbf{z}}_x$ where treatment $x$ was administered to the individual (whose $U=u$ and $\textbf{Z}=\textbf{z}$ are fixed over time) and fit a linear model with the outcome measurement $Y_{k^{\textbf{z}}_x}$ as the dependent variable and time $k^{\textbf{z}}_x$ or a function of time as the independent variable. If we reject the hypothesis of no association between time and the outcome, then we have evidence against stationarity \eqref{eq:stationarity1}, and therefore against the assumed outcome generating process \eqref{eq:Y1}. We can reject, for example, when carryover effects exist or other factors whose distribution changes over time affect some outcome measurements. Yet, this simple hypothesis test can have low power, and failure to reject the hypothesis of no association between time and outcome should not be interpreted as confirmation that the basic causal model holds true.

The assumptions of constant $f^Y$ and $F_\varepsilon$ over time and of no carryover effect are plausible only in specific circumstances. In work targeted towards practitioners, N-of-1 trials have been suggested to be ideal study designs in research on symptom management, palliative care, chronic non-malignant pain, and sleep disorders \cite{Essentials_ch4}. Indeed, N-of-1 trials are considered to be particularly suitable when the individual is affected by a chronic disease with “minimal fluctuations over time” of the symptoms, and the candidate treatment is an intervention that only has a transient effect on the symptoms (rapid onset of effect, quick reversal when the treatment is stopped, and no cumulative effect) without altering the underlying pathology \cite{Essentials_ch4}. These are the conditions under which Equation \eqref{eq:Y1} is plausible. We will show that under Equation \eqref{eq:Y1}, estimating meaningful causal estimands in N-of-1 trials is simple.

\section{Causal estimands} \label{sec:causal_estimands}

We focus on effects on the additive scale comparing always treated to never treated strategies, but our results can be easily adapted to different scales and contrasts.

We define the individual causal effect for individual $i$ at time $k$ as
\begin{equation}
\label{eq:ice}
    ICE_{i,k}=Y^{\bar{a}_k = \bar{1}_k}_{i,k} - Y^{\bar{a}_k = \bar{0}_k}_{i,k}
\end{equation}
with $\bar{1}_k$ being a vector of size $k$ with elements equal to $1$, and $\bar{0}_k$ defined similarly. This estimand represents the contrast between the counterfactual outcome at time $k$ under always treatment strategy and the counterfactual outcome at the same time under the never treated strategy. We define this causal estimand as the individual causal effect to be consistent with the definitions commonly used in epidemiology, where this definition is often attributed to $ICE_{i,1}$ \cite{hernan2019}. It is not possible to observe both counterfactuals at time $1$ (or any other time) for the same individual, because an individual either received the treatment at this time or they did not \cite{hernan2019}. Indeed, the $ICE_{i,k}$ is only identifiable under strong assumptions.

Yet, when we conduct an N-of-1 trial, we are restricting our inference to a single individual's characteristics. We can interpret this as conditioning on the characteristics of the individual $i$ ($U=u_i$) at the beginning of the study. We call $U\text{-}CATE$ the average treatment effect conditional on the individual-specific variable $U$.
More formally, we define the average treatment effect at time $k$ conditioning on $U=u$ as 
\begin{equation} \label{eq:ucate1}
    U\text{-}CATE_k(u) = \mathbb{E}(Y^{\bar{a}_k = \bar{1}_k}_{k} \mid U=u) - \mathbb{E}(Y^{\bar{a}_k = \bar{0}_k}_{k} \mid U=u).
\end{equation}
The definition of this estimand follows the same rationale of the estimand in Robins et al.\ \cite{Robins1999}. The $ U\text{-}CATE_k(u)$ represents the contrast between the expected counterfactual outcome at time $k$, had an individual with $U=u$ always received treatment until time $k$, and the expected counterfactual outcome at time $k$ had an individual with $U=u$ never received treatment until time $k$. We argue that $ U\text{-}CATE_{k}(u)$ is a relevant estimand to consider in N-of-1 trials, being the conditional version of the conventionally targeted causal estimand in population-level studies with time-varying exposure \cite{Robins1999, hernan2019}.

The $U\text{-}CATE_k(u)$ is a conditional average treatment effect \cite{hernan2019}. It is in principle observable in an experiment where many individuals with the same characteristics $U=u$ are randomized to either receive treatment or control at all time points. In such an experiment, it would also be possible to directly assess the stationarity conditions in \eqref{eq:stationarity1} and \eqref{eq:stationarity1_expectation}. However, since $U$ is likely high dimensional, two individuals are unlikely to share the same value of $U$, implying that this estimand can therefore be informally interpreted as a causal effect specific to the individual. Indeed, consistent with previous work \cite{Daza2019n1rt, Robins1999}, we refer to $U\text{-}CATE_k(u)$ as an “individual-specific causal effect”. 

To make explicit the relationship between the estimands of N-of-1 trials and parallel group RCTs, we also define a “population-level causal effect”, which is more common in medicine and epidemiology, 
\begin{equation}
\label{eq:ACE}
    ACE_k = \mathbb{E}(Y^{\bar{a}_k = \bar{1}_k}_{k}) - \mathbb{E}( Y^{\bar{a}_k = \bar{0}_k}_{k} ),
\end{equation}
which is a contrast between the expected value of the counterfactual outcome at time $k$ under the always treated strategy and under the never treated strategy in some population of interest (including subjects with different values of $U$). 
The $ACE_k$, as defined above, is a natural causal estimand in an RCT that randomizes individuals to either taking the treatment at every time until $k$ or to not taking the treatment through time $k$.

\section{Inference from N-of-1 trials under the “basic scenario”} 
\label{sec:inference_basic_scenario}

In this section, we discuss identification and estimation, in an N-of-1 trial, of causal estimands presented in Section \ref{sec:causal_estimands}, assuming the basic data generating process described in Section \ref{sec:basic_scenario}. 

In our setting, the treatment schedule $\textbf{Z}$ can be interpreted as an ancillary statistic \cite{hernan2019, Robins2000}, analogous to the flip of a coin in the example from Robins and Wasserman \cite{Robins2000}. The causal model described in Section \ref{sec:basic_scenario} can be interpreted, in fact, as a “mixed experiment” \cite{Robins2000}. The conditionality principle \cite{hernan2019, Robins2000} asserts that the inference we make when $\textbf{Z}=\textbf{z}$ is randomly assigned should be identical to the inference we make when $\textbf{Z}=\textbf{z}$ is assigned without randomization. In accordance with this principle, we conduct our inference conditional on $\textbf{Z}$ (see Section \ref{sec:randomization} for further discussion).

Under the causal model described in equations \eqref{eq:A1}, \eqref{eq:Y1}, and condition \eqref{eq:Z1}, we have that $U\text{-}CATE_{k}(u)$ is constant over time, see Appendix A. Moreover, the $U\text{-}CATE_{k}(u)$ is identified by
\begin{equation} \label{eq:tau1}
    \tau(u,\textbf{z}) = \mathbb{E}(Y_{k^{\textbf{z}}_1} \mid A_{k^{\textbf{z}}_1}=1, U=u, \textbf{Z}=\textbf{z}) - \mathbb{E}(Y_{k^{\textbf{z}}_0} \mid A_{k^{\textbf{z}}_0}=0, U=u, \textbf{Z}=\textbf{z})
\end{equation}
for any $k^{\textbf{z}}_1 \in \{k_* \in \{1,...,t\} \mid f^{A}(\textbf{z},k_*)=1\}$ and $k^{\textbf{z}}_0 \in \{k_* \in \{1,...,t\} \mid f^{A}(\textbf{z},k_*)=0\}$ (see Appendix A). This quantity corresponds to the difference between the expected value of the outcome for an individual with characteristics $U=u$ and treatment schedule $\textbf{Z}=\textbf{z}$ when the individual is treated and the same expected value when no treatment is administered. 

In describing the data generating process using equations \eqref{eq:A1} and \eqref{eq:Y1} we committed to a “non-parametric structural equation model with independent errors” (NPSEM-IE) \cite{pearl2009}. These models implicitly assume several untestable cross-world independencies between counterfactuals \cite{hernan2019}. We emphasize that our identification result does not depend on such cross-world independencies and it can be obtained also assuming only condition \eqref{eq:Z1}, consistency \eqref{eq:consistency}, no carryover \eqref{eq:carryov1}, the causal DAG in Figure \ref{fig:dgm1} under a finest fully randomized causally interpreted structured tree graph (FFRCISTG) model \cite{Robins1986} (which includes the NPSEM-IE as a strict submodel \cite{Richardson2013}), and the stationarity of the expected conditional counterfactual outcome \eqref{eq:stationarity1_expectation}, see Appendix A. Since we consider a causal effect on the additive scale, strong stationarity at the level of the expected value \eqref{eq:stationarity1_expectation} is sufficient for identification, without requiring the stricter assumption \eqref{eq:stationarity1} of strong stationarity of the distribution. 

Denote the difference between the mean of the observed outcomes of the individual $i$ during the time points where treatment is received and the mean of the observed outcomes during the time points where treatment is not received by
\begin{equation}
\label{eq:tauhat1}
\begin{split}
&\hat{\tau_i} = \\
&\frac{1}{\sum^{t}_{k=1} \{ \mathbb{I}(A_{i,k}=1) \}} \sum^{t}_{k=1} \{ \mathbb{I}(A_{i,k}=1) Y_{i,k} \} - \frac{1}{\sum^{t}_{k=1} \{ \mathbb{I}(A_{i,k}=0) \}} \sum^{t}_{k=1} \{ \mathbb{I}(A_{i,k}=0) Y_{i,k} \}.
\end{split}
\end{equation}
This quantity is a function of data from a single individual $i$, as would be available in an N-of-1 trial. We know that $\hat{\tau_i}$ is well defined because condition \eqref{eq:Z1} guarantees that both denominators are non-zero for the treatment schedule $\textbf{z}_i$. 

We can show that $\hat{\tau_i}$ is an unbiased estimator for $\tau(u_i,\textbf{z}_i)$ conditional on $U=u_i$ and $\textbf{Z}=\textbf{z}_i$, see Appendix B. 
This means that if we were to repeat the N-of-1 experiment with treatment schedule $\textbf{z}_i$ for infinite individuals with $U=u_i$, the average of the estimates obtained according to \eqref{eq:tauhat1} across the repetitions would correspond to the individual-specific causal effect $\tau(u_i, \textbf{z}_i)$. We emphasize that, in our statistical inference framework, we require conceptualization of a superpopulation of individuals with $U=u_i$, though in practice few or no individuals will share all characteristics represented by $U$.

The estimator $\hat{\tau_i}$ converges towards $\tau(u_i, \textbf{z}_i)$ as the duration of the experiment increases (as $t$ goes to infinity) for an individual with characteristics $U=u_i$ and treatment schedule $\textbf{Z}=\textbf{z}_i$. Conditional on the context $U=u_i, \textbf{Z}=\textbf{z}_i$, the estimator distribution can be approximated by a normal distribution with mean $\tau(u_i,\textbf{z}_i)$ when $t$ is large,
\begin{equation}
\label{eq:basic_z_distribution}
\hat{\tau_i} \approx \mathcal{N}\bigg(\tau(u_i,\textbf{z}_i), \frac{\sigma^2(1,u_i)}{\sum^{t}_{k=1} \{ \mathbb{I}(f^{A}(\textbf{z}_i,k)=1) \}} + \frac{\sigma^2(0,u_i)}{\sum^{t}_{k=1} \{ \mathbb{I}(f^{A}(\textbf{z}_i,k)=0) \}} \bigg),
\end{equation}
see Appendix B for the derivation.

This result enables construction of simple statistical tests for the null hypothesis of no individual-specific causal effect. Specifically, for large $t$, it justifies the use of an unpaired z-test or an unpaired t-test of the null hypothesis that $U\text{-}CATE_k(u_i)=0$. The unpaired t-test has been described as the simplest way to test for an effect in an N-of-1 trial, and is considered an appropriate test when carryover or other more complex relationships are absent \cite{DEcIDE}. Our results justify this claim when the $U\text{-}CATE_k(u_i)$ is the causal estimand of interest. The variance in \eqref{eq:basic_z_distribution} is also consistent with informal claims in the applied literature on single-case experiments. In particular, it has been claimed that if the symptoms of the disease under investigation fluctuate substantially over time, it is still possible to conduct meaningful single-case studies, but the study will have to last longer \cite{Essentials_ch15}. A substantial fluctuation of the symptoms is represented by high $\sigma^2(1,u_i)$ and $\sigma^2(0,u_i)$, which leads to an imprecise estimator of the effect if it is not compensated by a long follow-up period. In practice, $\sigma^2(1,u_i)$ and $\sigma^2(0,u_i)$ can be estimated by applying the unbiased sample variance estimator to the outcome measurements obtained from the individual $i$ when they were treated and not treated, respectively.

Under the additional assumption of equality of the outcome conditional variances under both treatments
\begin{equation} \label{eq:equal_sigma}
  \sigma^2(1,u_i)=\sigma^2(0,u_i)=\sigma^2(u_i),  
\end{equation} the estimator's conditional variance can be approximated by
\begin{equation} \label{eq:approx_variance}
     Var(\hat{\tau_i} \mid U=u_i, \textbf{Z}=\textbf{z}_i) \approx \frac{\sigma^2(u_i)}{t \cdot \alpha(1-\alpha)},
\end{equation}
with $\alpha$ being the proportion of treated time points in the cycle of length $q$ when $\textbf{Z}=\textbf{z}_i$, see Appendix B for the derivation. Assumption \eqref{eq:equal_sigma} always holds when $f^Y$ is an additive noise model \cite{Peters2013} or under the hypothesis that the treatment has no effect on the conditional outcome distribution, i.e., $\mathbb{P}(Y^{a_k=1}_{k}=y \mid U=u_i) = \mathbb{P}(Y^{a_k=0}_{k}=y \mid U=u_i)$. Here, $\sigma^2(u_i)$ can be estimated using the conventional pooled variance estimator \cite{Ruxton2006}.
Equation \eqref{eq:approx_variance} makes explicit that, under the considered causal model and assumption \eqref{eq:equal_sigma}, balanced treatment schedules with the same number of treated and untreated time points (i.e., $\alpha=0.5$) lead to a more precise estimation of the individual-specific causal effect.

In Appendix C, we discuss the setting where $\varepsilon_{i,k}$ is constant in $i$ and $k$, i.e., the “error" terms are the same for every unit and at every point in time, and, as a consequence $\sigma^2(1,u_i)$ and $\sigma^2(0,u_i)$ are equal to zero. In this extreme scenario, $ICE_{i,k}$ is constant over time and $U\text{-}CATE_k(u_i)=ICE_{i,k}$. Then, the estimator $\hat{\tau_i}$ would correspond to the individual causal effect $ICE_{i,k}$. However, an investigator may be reluctant to assume that $\varepsilon_{i,k}$ is fixed, as this implies that the common causes $U$, instantiated at the beginning of the trial, and the assigned treatment are sufficient for the perfect prediction of the outcome measurements during the full study period. The assumption of constant counterfactuals was used by Hernan and Robins \cite{hernan2019} in their didactic example to demonstrate the identification of the individual causal effect in a cross-over experiment with one unit and two time points. 
The assumption of constant $\varepsilon_{i,k}$ is falsifiable in an N-of-1 trial with several time points. Indeed, if $\varepsilon_{i,k}$ is constant over time, the outcome measurement for an individual should always appear exactly identical for the same level of the treatment status, that is, $Y_{i,k}-Y_{i,k'}=0$ for all $k, k'$ such that $A_{i,k}=A_{i,k'}$.

\section{Complex scenarios} \label{sec:complex_scenarios}

The key feature of the “basic scenario" data generating process that enabled the simple estimators derived in Section \ref{sec:inference_basic_scenario} is the strong conditional stationarity captured by Equation \eqref{eq:stationarity1}. In this section, we will discuss threats to strong stationarity, propose a relaxed causal model imposing weaker stationarity restrictions that mitigate those threats, and describe how to estimate causal effects under the relaxed causal model.

\subsection{Threats to strong stationarity described in the N-of-1 trials literature} \label{sec:stationarity}

Stationarity is a common assumption in the N-of-1 trials literature \cite{Daza2018-lo, daza2019, daza2022, neto2016} and in the field of causal analysis of time series more broadly \cite{Runge2023, Eichler2009}. There are several types of stationarity assumptions, and all require some form of regularity over time of the time series' probabilistic behavior \cite{neto2016}.
While in parallel-group studies we observe a sample of independent observations from the same distribution, in an N-of-1 trial we only observe one instance of the time series \cite{neto2016}. Conceptually, assuming some form of stationarity allows us to think about this one-instance sequence of data as a sample of independent observations from some time invariant distributions, even if the data is about only one individual \cite{neto2016}.

In the N-of-1 trials literature, common reasons for the violation of stationarity assumptions are extensively described, both informally --- from a clinical perspective --- and in statistical terms. Specifically, strong correlation between outcome measurements for the same individual \cite{Schork2018-xs, Lillie2011-gv, Schmid2022-oh, Essentials_ch6}, carryover effects \cite{Schork2018-xs, Lillie2011-gv, Schmid2022-oh, Essentials_ch6}, and time trends \cite{Schmid2022-oh} are considered to be the main challenges when designing and analyzing N-of-1 trials.
As already mentioned in Section \ref{sec:counterfactual_outcomes}, “carryover effects” are present when the effect of a treatment administered at one time period carries over to outcomes in subsequent periods \cite{Essentials_ch6}. In other words, there is a carryover effect when the treatment does not stop exerting effects on the outcome (directly or indirectly) after the interval of administration, see Equation \eqref{eq:carryov1}. Carryover effects are considered a serious problem in N-of-1 trials, and in some circumstances the design is deemed unfeasible due to their presence \cite{Essentials_ch6}. Consider for example an N-of-1 trial that aims to compare the effects of two vaccines on incidence of infection. If the vaccines induce life-long immunity, and the individual is first treated with the first vaccine, no information can be available about what would have happened if the individual was treated with the second vaccine instead.

The correlation between outcome measurements for the same individual, sometimes called “autocorrelation”, refers to a statistical, and not necessarily a causal, property. Autocorrelation is sometimes attributed to the existence of a common cause, other than the treatment and possibly unmeasured, of a proper subset of outcome measurements for the individual \cite{Schmid2022-oh}. In Section \ref{sec:relaxed_scenario} we will discuss the scenario in which such common causes exist and are measured.
Autocorrelation is also sometimes attributed to the presence of a direct causal effect of the outcome state at one time point on outcome states at later time points \cite{Daza2018-lo, daza2022}. We will refer to this particular source of outcome correlation as an “outcome-outcome effect” and formalize it in terms of structural equations (see Equation \eqref{eq:Y2}) to avoid confusion with the statistical terminology. 

The term “time trend” has also been used in an inconsistent way in the N-of-1 trials literature. Informally, this phenomenon has been interpreted as the existence of a time-varying variable, other than the treatment, that causes the outcome and whose distribution changes over time (see Section \ref{sec:time_trend} for more details). 

The presence of common causes of some outcome measurements and the issue of outcome-outcome effects may be related to that of carryover. Indeed, if the treatment has an effect on the outcome, the presence of outcome-outcome effects ensures that the treatment at time $k$ has an indirect effect through the outcome $Y_k$ on subsequent outcome measurements $Y_{s}$ for $s> k$. Similarly, if the treatment has an effect on a common cause of two outcome measurements, carryover will be present. Furthermore, the concept of time trend is also related to the presence of common causes of some outcome measurements. Indeed, if the time-varying variable at time $k$ that causes the outcome at time $k$ affects future realization of itself, it represents a common cause of outcome measurements.

The terms “autocorrelation”, “time trend”, and “carryover” are often used informally and their interpretation is ambiguous if the full causal model is not carefully specified. We use these terms simply to draw connections to the N-of-1 trials literature, but warn the reader that these terms do not indicate distinct clearly defined concepts.

The basic causal model described in Section \ref{sec:basic_scenario} forbids carryover effects, outcome-outcome effects, and time-varying causes of the outcome whose distribution changes over time or that affect multiple outcome measurements. In the next sections, we will discuss identification and estimation of causal effects when these phenomena may be present. While these phenomena do indeed lead to violations of strong conditional stationarity \eqref{eq:stationarity1}, as noted above, they can be compatible with relaxed notions of stationarity under which it is still possible to estimate individual-specific effects from N-of-1 trials, so long as certain key variables are observed.

\subsection{A relaxed causal model} \label{sec:relaxed_scenario}

Consider the same treatment assignment as in Equation \eqref{eq:A1}. Suppose that, additionally, at every time $k$ we have measurements of a variable $L_k$, generated according to the structural equation
\begin{equation} \label{eq:L1}
    L_{k} = f^L(A_{k}, Y_{k-1}, L_{k-1}, U_L, \gamma_{k})
\end{equation}
with $\gamma_{k}$ being an exogenous random variable with distribution $F_{\gamma}$ that does not change over time, and $f^L$ being a function that does not depend on $k$. We assume that $\bar{\gamma}$ is a vector of i.i.d. exogenous random variables. The variable $U_L$ represents the set of common direct causes of all covariate measurements during the study period, instantiated at baseline. Equation \eqref{eq:L1} introduces in our causal model a covariate $L_k$ that is affected by its previous state $L_{k-1}$, the treatment status $A_k$, the previous state of the outcome $Y_{k-1}$, and $U_L$.

The outcome $Y_{k}$ at time $k$ is now generated by a structural equation,
\begin{equation} \label{eq:Y2}
Y_{k} = f^Y(L_{k}, A_{k}, Y_{k-1}, L_{k-1}, A_{k-1}, U_Y, \varepsilon_{k}).\\
\end{equation}
As in Section \ref{sec:basic_scenario}, $\bar{\varepsilon}$ is a vector of i.i.d.\ exogenous variables drawn from a distribution $F_{\varepsilon}$ that does not vary over time. The variable $U_Y$ represents the common direct causes of all the outcome measurements, instantiated at baseline. 

For didactic purposes, we describe the setting where $Y_k, L_k, U_L, U_Y$ are discrete with finite support, and where $Y_k$ and $L_k$ are directly affected by time-varying variables instantiated only at time $k$ and $k-1$, but our results can be generalized to the continuous case and to settings where variables have direct effects over longer time horizons. Moreover, for simplicity, we will assume that at time 1, both $L_1$ and $Y_1$ are generated according to Equations \eqref{eq:L1} and \eqref{eq:Y2}. Specifically, we will assume that the variables $A_0 \text{ (or }f^{A}(\textbf{Z},0)), L_0$ and $Y_0$, which do not exist or are not measured, are fixed to constant values $a^*_0, l^*_0, y^*_0$. In practice, a researcher may want to consider the study period as beginning at time 2, rather than time 1, and consider time point 1 as “fixed”.

As before, we will assume causal consistency
\begin{equation} \label{eq:consistency2}
  Y_{k} = Y^{\bar{a}_{k}=\bar{v}_{k}}_{k}, L_{k} = L^{\bar{a}_{k}=\bar{v}_{k}}_{k} \text{ when } \bar{A}_{k}=\bar{v}_{k}
\end{equation}
for all $k$ and $\bar{v}_{k} \in \{0,1\}^k$.

We will denote with $U$ the union of $U_L$ and $U_Y$. As in the basic scenario (Section \ref{sec:basic_scenario}), the variable $U$ includes the common causes of all the outcome measurements, either directly or indirectly (through $\bar{L}$). Therefore, this variable can still be interpreted as an individual-specific variable, potentially high-dimensional, summarizing characteristics of the individual at baseline and representing the source of effect heterogeneity. However, differently from the interpretation in the basic scenario (Section \ref{sec:interpretation_causal_model}), $U$ can now include some components that do not affect all the outcome measurements $\bar{Y}$ directly, as long as they affect all covariate measurements $\bar{L}$. 

We will restrict the possible treatment schedules in such a way that each individual, at least once during a cycle, receives the comparator and the treatment twice in a row
\begin{equation} \label{eq:Z2}
\mathcal{Z} \subseteq \bigg\{ \textbf{z}=(z_1,...,z_q) \bigg\rvert \sum^{q-1}_{k=1} \mathbb{I}(z_{k} = x) \mathbb{I}(z_{k+1} = x) >0, \, \forall x \in \{ 0,1 \} \bigg\}.
\end{equation}

Furthermore, for simplicity, we will require that every possible trajectory of the time-varying variables has a non-zero probability of being observed under any intervention strategy. Therefore, our positivity requirement can be written as
\begin{equation} \label{eq:positivity}
    \mathbb{P}(\bar{Y}^{\bar{a}=\bar{v}}=\bar{y}, \bar{L}^{\bar{a}=\bar{v}}=\bar{l} \mid U=u) > 0
\end{equation}
for every $\bar{y} \in \mathcal{Y}^t$, every $\bar{l} \in \mathcal{L}^t$, every $u \in \mathcal{U}$, and every $\bar{v} \in \{0,1\}^t$, where $\mathcal{L}$ and $\mathcal{Y}$ represent the sets of possible values of the time-varying covariate and the outcome, respectively. In practice, this positivity assumption can be relaxed to only concern trajectories compatible with the structural equations, excluding for example trajectories that are impossible due to determinisms.

\begin{figure}
    \begin{minipage}{1\textwidth}
        \centering
                \begin{tikzpicture}
                    \tikzset{line width=1.5pt, outer sep=0pt,
                    ell/.style={draw,fill=white, inner sep=4pt,
                    line width=1.5pt}};

                    \node[name=Z,ell,  shape=ellipse] at (0,0) {$\textbf{Z}$};
                    \node[name=U,ell,  shape=ellipse, dashed] at (0,6) {$U$};
                    \node[name=A_1,ell,  shape=ellipse] at (3,0) {$A_{1}$};
                    \node[name=Y_1,ell,  shape=ellipse] at (5,2) {$Y_{1}$};
                    \node[name=L_1,ell,  shape=ellipse] at (3.5,4.7) {$L_{1}$};
                    \node[name=A_2,ell,  shape=ellipse] at (6,0) {$A_{2}$};
                    \node[name=Y_2,ell,  shape=ellipse] at (8,2) {$Y_{2}$};
                    \node[name=L_2,ell,  shape=ellipse] at (6.5,4.7) {$L_{2}$};
                    \node[name=spa, draw=none] at (10.5,0) {...};
                    \node[name=spa2, draw=none] at (12.5,2) {...};
                    \node[name=spa3, draw=none] at (11,4.7) {...};
                    \node[name=A_3,ell,  shape=ellipse] at (9,0) {$A_{3}$};
                    \node[name=Y_3,ell,  shape=ellipse] at (11,2) {$Y_{3}$};
                    \node[name=L_3,ell,  shape=ellipse] at (9.5,4.7) {$L_{3}$};

                     \begin{scope}[>={Stealth[black]},
                                  every edge/.style={draw=black,very thick}]
                        \path [->] (Z) edge[color=green] (A_1); 
                        \path [->] (Z) edge[color=green, bend right=20] (A_2);
                        \path [->] (Z) edge[color=green, bend right=30] (A_3); 
                        \path [->] (U) edge (Y_1);
                        \path [->] (U) edge[bend left=25] (Y_2);
                        \path [->] (U) edge[bend left=32] (Y_3); 
                        \path [->] (U) edge (L_1);
                        \path [->] (U) edge[bend left=15] (L_2);
                        \path [->] (U) edge[bend left=20] (L_3); 
                        \path [->] (A_1) edge (L_1);
                        \path [->] (A_1) edge (Y_1);
                        \path [->] (A_1) edge (Y_2);
                        \path [->] (A_2) edge (L_2);
                        \path [->] (A_2) edge (Y_2);
                        \path [->] (A_2) edge (Y_3);
                        \path [->] (A_3) edge (L_3);
                        \path [->] (A_3) edge (Y_3);
                        \path [->] (L_1) edge (L_2);
                        \path [->] (L_1) edge (Y_1);
                        \path [->] (L_1) edge (Y_2);
                        \path [->] (L_2) edge (Y_2);
                        \path [->] (L_2) edge (Y_3);
                        \path [->] (L_3) edge (Y_3);
                        \path [->] (L_2) edge (L_3);
                        \path [->] (Y_1) edge (L_2);
                        \path [->] (Y_1) edge (Y_2);
                        \path [->] (Y_2) edge (L_3);
                        \path [->] (Y_2) edge (Y_3);
                    \end{scope}
                \end{tikzpicture}
\end{minipage}
    \caption{Directed Acyclic Graph representing the causal relationships between the assigned treatment schedule ($\textbf{Z}$), the received exposure ($A_k$), the outcome ($Y_k$), the individual-specific random variable ($U$) and the time-varying covariate ($L_k$) in the relaxed causal model for $t>2$. Nodes represent variables, directed edges represent causal relationships. Dashed nodes represent unobserved variables and green edges represent deterministic causal relationships. To avoid clutter, we do not depict explicitly the exogenous noise random variables $\varepsilon_k$ and $\gamma_k$ that affect respectively $Y_k$ and $L_k$.}
    \label{fig:dgm3}
\end{figure}
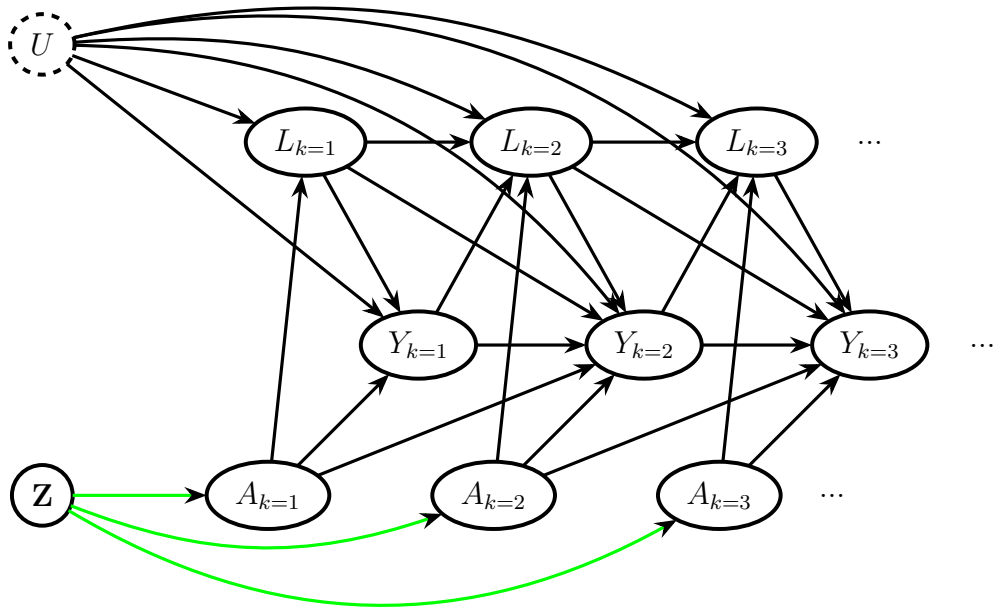

The causal model described by Equations \eqref{eq:A1}, \eqref{eq:L1}, and \eqref{eq:Y2} is represented in the DAG in Figure \ref{fig:dgm3}. As before, the DAG cannot represent the non-graphical assumptions, such as the requirement that the structural equations or the random noise distributions do not change across time.

In this new causal model, the treatment variables $\bar{A}$ are generated exactly as in the basic scenario in Section \ref{sec:treatment_assignment}; the outcome variable at time $k$ is a function of the treatment assigned at time $k$, an individual-specific variable $U_Y$, and an independent noise variable $\varepsilon_k$. However, in this new causal model, $Y_k$ is also determined by its previous state $Y_{k-1}$, therefore allowing for outcome-outcome effects. More generally, we are allowing for the presence of carryover effects, both directly and indirectly (through $L_k$ and $Y_k$). Indeed, the treatment status at time $k$ ($A_k$) has a direct effect on the outcome at the next time ($Y_{k+1}$), and an indirect effect on all the subsequent outcome measurements through its effect on $L_k$ and $Y_k$. Furthermore, the time-varying covariate $L_k$ represents a common cause of a proper subset of outcome measurements when $k>1$, and could induce a time trend in the outcome, since its distribution changes over time (see Section \ref{sec:time_trend} for more details). Indeed, every informal threat to strong stationarity discussed in Section \ref{sec:stationarity} is present, and the assumption of strong stationarity in Equation \eqref{eq:stationarity1} does not generally hold under this causal model. As a consequence, the individual-specific causal effect $U\text{-}CATE_k(u)$ may not be constant over time, but rather changes based on the value of $k$. 

The causal model nevertheless imposes regularity constraints over time: the function generating the variable $L_k$ ($f^L$) does not change over time; the noise variable distributions $F_{\gamma}$ remain the same over time; and time-varying variables can only have a direct effect on variables measured in the same time point or the next time point. These assumptions derived from \eqref{eq:L1} imply that
\begin{equation} \begin{split} \label{eq:stationarityL}
\mathbb{P}(L^{\bar{a}_k=(\bar{v}, x)}_{k}=l_2 \mid Y^{\bar{a}_{k-1}=\bar{v}}_{k-1}=y_1, L^{\bar{a}_{k-1}=\bar{v}}_{k-1}=l_1, U=u) = \\
\mathbb{P}(L^{\bar{a}_{k'}=(\bar{v}', x)}_{k'}=l_2 \mid Y^{\bar{a}_{k'-1}=\bar{v}'}_{k'-1}=y_1, L^{\bar{a}_{k'-1}=\bar{v}'}_{k'-1}=l_1, U=u)
\end{split}
\end{equation}
for every $k, k' \in \{2,...,t\}$, for every $l_2,l_1 \in \mathcal{L}$, for every $y_1 \in \mathcal{Y}$, for every $u \in \mathcal{U}$, for every vector $\bar{v} \in \{0,1\}^{k-1}, \bar{v}' \in \{0,1\}^{k'-1}$ and $x \in \{0,1\}$. Stationarity \eqref{eq:stationarityL} also applies to the first time point, provided that  $y_1=y^*_0$ and $l_1=l^*_0$.

Similarly, our non-graphical assumptions about the stability over time of $f^Y$ and $F_\varepsilon$ induce a relaxed version of stationarity for the outcome. Indeed, we have that 
\begin{equation} 
\begin{split}
\label{eq:stationarity2}
&\mathbb{P}(Y^{\bar{a}_k=(\bar{v}, x_1,x_2)}_{k}=y_2 \mid L^{\bar{a}_k=(\bar{v},x_1,x_2)}_{k}=l_2, Y^{\bar{a}_{k-1}=(\bar{v},x_1)}_{k-1}=y_1, L^{\bar{a}_{k-1}=(\bar{v},x_1)}_{k-1}=l_1, U=u) = \\
&\mathbb{P}(Y^{\bar{a}_{k'}=(\bar{v}', x_1,x_2)}_{k'}=y_2 \mid L^{\bar{a}_{k'}=(\bar{v}',x_1,x_2)}_{k'}=l_2, Y^{\bar{a}_{k'-1}=(\bar{v}',x_1)}_{k'-1}=y_1, L^{\bar{a}_{k'-1}=(\bar{v}',x_1)}_{k'-1}=l_1, U=u)
\end{split}
\end{equation}
for every $k, k' \in \{2,...,t\}$, for every $l_2,l_1 \in \mathcal{L}$, for every $y_2,y_1 \in \mathcal{Y}$, for every $u \in \mathcal{U}$, for every vector $\bar{v} \in \{0,1\}^{k-2}, \bar{v}' \in \{0,1\}^{k'-2}$ and $x_1,x_2 \in \{0,1\}$. The outcome transition at the first time point is also described by the same conditional distribution as the outcome transitions at subsequent time points, i.e., $\mathbb{P}(Y^{a_1=x_2}_{1}=y_2 \mid L^{a_1=x_2}_{1}=l_2, Y_0=y^*_0, L_0=l^*_0, A_0=a^*_0, U=u)$ equals the quantities in \eqref{eq:stationarity2} with $x_1=a^*_0, l_1=l^*_0, y_1=y^*_0$.

Informally, the stationarity assumption in \eqref{eq:stationarity2} allows us to consider, once again, the time series of outcome measurements as i.i.d.\ samples from fixed conditional distributions, despite now we condition on a larger set of variables.

\section{Inference from N-of-1 trials under the “relaxed causal model”} 
\label{sec:inference_relaxed_model}

We define
\begin{equation*}
\begin{split}
&k^{\textbf{z}}_{x} \in \{k_* \in \{1,...,t\} \mid f^{A}(\textbf{z},k_*)=x\}\\
&k^{\textbf{z}}_{x_1x_2} \in \{k_* \in \{1,...,t\} \mid \mathbb{I}(f^{A}(\textbf{z},k_*)=x_2)\mathbb{I}(f^{A}(\textbf{z},k_*-1)=x_1)=1\}
\end{split}
\end{equation*}
as generic time points in which a certain sequence of treatments is observed under the treatment schedule $\textbf{Z}=\textbf{z}$. For example, for a specific treatment schedule $\textbf{z}$, $k^{\textbf{z}}_{00}$ represents a time where no treatment was administered twice in a row, and $k^{\textbf{z}}_1$ represents a time where treatment was administered.
We will also define
\begin{equation}
\label{eq:gL}
g_{k^{\textbf{z}}_{x}}^L(l \mid y', l', u) =  \mathbb{P}(L_{k^{\textbf{z}}_{x}}=l \mid A_{k^{\textbf{z}}_{x}}=x, Y_{k^{\textbf{z}}_{x}-1}=y', L_{k^{\textbf{z}}_{x}-1}=l', U=u, \textbf{Z}=\textbf{z})
\end{equation}
as the distribution of the covariate at a time when treatment $x$ was administered, conditional on all its direct causes. Due to the stationarity condition \eqref{eq:stationarityL}, these conditional probabilities are the same for all time points $k^{\textbf{z}}_{x}$ in which treatment $x$ is administered. Similarly, define the conditional distribution 
\begin{equation}
\label{eq:gY}
\begin{split}
&g_{k^{\textbf{z}}_{x_1x_2}}^Y(y \mid l, y', l', u) = \\
&\mathbb{P}(Y_{k^{\textbf{z}}_{x_1x_2}}= y \mid L_{k^{\textbf{z}}_{x_1x_2}}=l, A_{k^{\textbf{z}}_{x_1x_2}}=x_2, Y_{k^{\textbf{z}}_{x_1x_2}-1}=y', L_{k^{\textbf{z}}_{x_1x_2}-1}=l', A_{k^{\textbf{z}}_{x_1x_2}-1}=x_1, U=u, \textbf{Z}=\textbf{z})
\end{split}
\end{equation}
for the outcome at a time when treatment $x_2$ has been assigned after treatment $x_1$, conditional on all outcome's direct causes. Due to the stationarity condition \eqref{eq:stationarity2}, these conditional probabilities are the same for all time points $k^{\textbf{z}}_{x_1x_2}$ in which the treatments $x_1,x_2$ are administered (see Appendix D).

Consider a hypothetical study of length $t>q$ where treatment is assigned according to \eqref{eq:A1}, time-varying covariates and outcomes are generated according to \eqref{eq:L1} and \eqref{eq:Y2}, and conditions \eqref{eq:Z2} and \eqref{eq:positivity} hold. Under this causal model, the expected value of the counterfactual outcome for the intervention $\bar{x}_k \in \{0,1\}^k$ at time $k$ conditional on $U=u$, can be identified by a longitudinal g-formula \cite{Robins1999, hernan2019}
\begin{equation*}
  \mathbb{E}(Y^{\bar{a}_k=\bar{x}_k}_k \mid U=u) = \theta_k(\bar{x}_k,u,\textbf{z}),
\end{equation*}
where 
\begin{equation} \label{eq:theta}
\theta_k(\bar{x}_k,u,\textbf{z}) =\sum_{\bar{y}_{k}, \bar{l}_k} y_k \,
\prod^{k}_{m=1} g_{k^{\textbf{z}}_{x_{m-1}x_{m}}}^Y(y_{m} \mid l_{m}, y_{m-1}, l_{m-1},  u) \, g_{k^{\textbf{z}}_{x_m}}^L(l_m \mid y_{m-1}, l_{m-1}, u).
\end{equation}
More details on the identification result can be found in Appendix D. Therefore, the individual-specific causal effect at time $k$ for the characteristics $U=u_i$ of the individual $i$ who followed the treatment schedule $\textbf{Z}=\textbf{z}_i$, can therefore be identified as
\begin{equation*}
  U\text{-}CATE_k(u_i)=\theta_k(\bar{1}_k,u_i,\textbf{z}_i) - \theta_k(\bar{0}_k,u_i,\textbf{z}_i).
\end{equation*}

Recall that variables $A_0, L_0$ and $Y_0$ are fixed to constant values $a^*_0, l^*_0, y^*_0$ under any intervention. The task of estimating the $U\text{-}CATE_k(u_i)$ in an N-of-1 trial amounts therefore to estimating the conditional probabilities 
\begin{enumerate}[label=\roman*), leftmargin=*]
\item $g_{k^{\textbf{z}_i}_{11}}^Y(y \mid l, y', l', u_i)$
\item $g_{k^{\textbf{z}_i}_{00}}^Y(y \mid l, y', l', u_i)$
\item $g_{k^{\textbf{z}_i}_{1}}^L(l \mid y', l', u_i)$
\item $g_{k^{\textbf{z}_i}_{0}}^L(l \mid y', l', u_i)$
\item $g_{k^{\textbf{z}_i}_{10}}^Y(y \mid l, y^*_0, l^*_0, u_i)$ if $a^*_0=1$ or $g_{k^{\textbf{z}_i}_{01}}^Y(y \mid l, y^*_0, l^*_0, u_i)$ if $a^*_0=0$
\end{enumerate}
for all possible values of $y,y' \in \mathcal{Y}$ and $l,l' \in \mathcal{L}$, using only data from the individual $i$. These conditional probabilities do not change over time. Therefore, for example, all time points in which treatment was administered twice in a row can be used to estimate i). Despite this, being able to estimate i), ii), iii), iv), or v) is a strong requirement, and it is not always possible to estimate these quantities well. Condition \eqref{eq:Z2} guarantees that the needed treatment sequences are observed at least once during the study period $t$, and condition \eqref{eq:positivity} guarantees a positive probability of observing any value of the time-varying variables at any time. However, it is not guaranteed that all the possible combinations of values are observed in a finite realization of $t$ measurements of the individual $i$. Therefore, estimating these conditional probabilities entirely non-parametrically, using the corresponding relative frequencies, may be unfeasible in practice. Introduction of parametric modeling assumptions may facilitate the estimation problem, smoothing over the unobserved combinations of the conditioning set. 

Suppose there exist estimators of i), ii), iii), iv), and v) that converge in probability towards the truth, for fixed $U=u_i$ and $\textbf{Z}=\textbf{z}_i$, as $t$ goes to infinity. Define $\hat{\theta}^{1}_{i,k}$ and $\hat{\theta}^{0}_{i,k}$ as the plug-in estimators obtained by replacing all terms in Equation \eqref{eq:theta} with estimates obtained by these estimators using data from the individual $i$ for the interventions $\bar{1}_k$ and $\bar{0}_k$, respectively \cite{hernan2019, Robins1999}. Then, the plug-in estimator $\hat{\theta}^{1}_{i,k}-\hat{\theta}^{0}_{i,k}$ converges towards $U\text{-}CATE_k(u_i)$, for fixed $U=u_i$ and $\textbf{Z}=\textbf{z}_i$, when $t$ goes to infinity \cite{hernan2019, Robins1999}. It is therefore possible to accurately estimate the individual-specific causal effect in an N-of-1 trial even in the relaxed causal model, provided that a sufficiently large number of time points are available and suitable estimators are used.

We emphasize that, while we assumed that treatment at a given time does not have a direct effect on the outcome beyond 1 time point in the future, we are allowing indirect carryover effects, through the outcome and the time-varying covariate, on \textit{all} future outcome measurements. This means that, under the specific restrictions implied by the causal model in Equations \eqref{eq:A1}, \eqref{eq:L1}, \eqref{eq:Y2}, \eqref{eq:Z2}, and \eqref{eq:positivity}, it is possible to estimate individual-specific causal effects of treatments with long-lasting effects in a single N-of-1 trial.

In our identification argument, we assumed $t>q$ only to guarantee that all four possible treatment combinations are observed. However, identification can also hold when $t=q$, provided that these combinations occur. Furthermore, in this section we considered for simplicity the setting where $Y_k$ and $L_k$ are only directly influenced by time-varying variables instantiated at time $k$ and $k-1$. By increasing this time window, we could represent more complex causal models, e.g., allowing for longer direct carryover effects. For example, $Y_k$ could also be affected directly by $A_{k-2}$ and $A_{k-3}$. It is straightforward to adapt our reasoning, our assumptions, and the g-formula in Equation \eqref{eq:theta} to accommodate these scenarios. Importantly, longer range direct treatment effects further restrict the possible values of $\textbf{Z}$ that allow nonparametric identification of the individual-specific causal effect, similarly to \eqref{eq:Z2}, by requiring that longer treatment sequences (e.g. $(0,0,0,0)$ and $(1,1,1,1)$) are assigned at least once. Our reasoning and assumptions can also be adapted to the case where the outcome and the time-varying variables are continuous, using well-established techniques \cite{hernan2019}. We emphasize that our identification result in \eqref{eq:theta} is indeed a special case of the g-formula for subject-specific effects derived by Robins et al.\ \cite{Robins1999}. To simplify the exposition, we motivated the result by assuming a NPSEM-IE model, but the identification result in \eqref{eq:theta} can be obtained assuming only consistency \eqref{eq:consistency2}, conditions \eqref{eq:Z2} and \eqref{eq:positivity}, the causal DAG in Figure \ref{fig:dgm3} under a FFRCISTG model \cite{Robins1986, Richardson2013}, and the stationarity assumptions \eqref{eq:stationarityL} and \eqref{eq:stationarity2}, see Appendix D. Furthermore, as shown in  Robins et al.\ \cite{Robins1999}, when there is a positive probability of receiving every possible treatment trajectory, stationarity of the counterfactuals (assumptions \eqref{eq:stationarityL} and \eqref{eq:stationarity2}) is not needed for identification, and a weaker stationarity assumption of the observed conditional distributions is sufficient for the estimation using data from only one individual, see additional discussion in Appendix D.

\subsection{The meaning of the term “time trend”} 
\label{sec:time_trend}

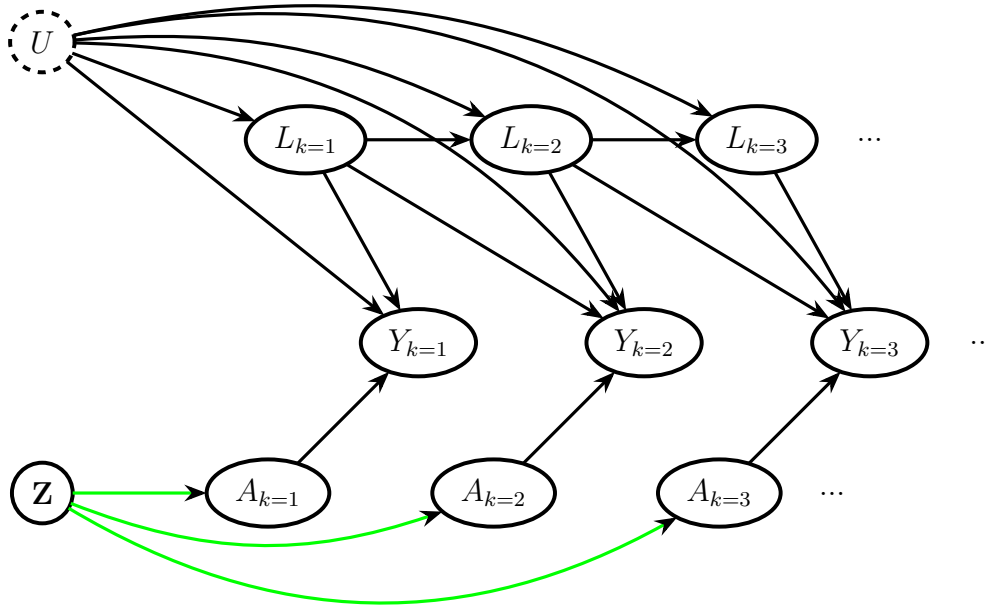
\begin{figure}
    \begin{minipage}{1\textwidth}
        \centering
                \begin{tikzpicture}
                    \tikzset{line width=1.5pt, outer sep=0pt,
                    ell/.style={draw,fill=white, inner sep=4pt,
                    line width=1.5pt}};

                    \node[name=Z,ell,  shape=ellipse] at (0,0) {$\textbf{Z}$};
                    \node[name=U,ell,  shape=ellipse, dashed] at (0,6) {$U$};
                    \node[name=A_1,ell,  shape=ellipse] at (3,0) {$A_{1}$};
                    \node[name=Y_1,ell,  shape=ellipse] at (5,2) {$Y_{1}$};
                    \node[name=L_1,ell,  shape=ellipse] at (3.5,4.7) {$L_{1}$};
                    \node[name=A_2,ell,  shape=ellipse] at (6,0) {$A_{2}$};
                    \node[name=Y_2,ell,  shape=ellipse] at (8,2) {$Y_{2}$};
                    \node[name=L_2,ell,  shape=ellipse] at (6.5,4.7) {$L_{2}$};
                    \node[name=spa, draw=none] at (10.5,0) {...};
                    \node[name=spa2, draw=none] at (12.5,2) {...};
                    \node[name=spa3, draw=none] at (11,4.7) {...};
                    \node[name=A_3,ell,  shape=ellipse] at (9,0) {$A_{3}$};
                    \node[name=Y_3,ell,  shape=ellipse] at (11,2) {$Y_{3}$};
                    \node[name=L_3,ell,  shape=ellipse] at (9.5,4.7) {$L_{3}$};

                     \begin{scope}[>={Stealth[black]},
                                  every edge/.style={draw=black,very thick}]
                        \path [->] (Z) edge[color=green] (A_1); 
                        \path [->] (Z) edge[color=green, bend right=20] (A_2);
                        \path [->] (Z) edge[color=green, bend right=30] (A_3); 
                        \path [->] (A_1) edge (Y_1);
                        \path [->] (A_2) edge (Y_2);
                        \path [->] (A_3) edge (Y_3);
                        \path [->] (U) edge (Y_1);
                        \path [->] (U) edge[bend left=25] (Y_2);
                        \path [->] (U) edge[bend left=32] (Y_3); 
                        \path [->] (U) edge (L_1);
                        \path [->] (U) edge[bend left=15] (L_2);
                        \path [->] (U) edge[bend left=20] (L_3); 
                        \path [->] (L_1) edge (Y_1);
                        \path [->] (L_1) edge (L_2);
                        \path [->] (L_2) edge (Y_2);
                        \path [->] (L_2) edge (L_3);
                        \path [->] (L_3) edge (Y_3);
                        \path [->] (L_1) edge (Y_2);
                        \path [->] (L_2) edge (Y_3);
                    \end{scope}
                \end{tikzpicture}
\end{minipage}
    \caption{Directed Acyclic Graph representing the causal relationships between the assigned treatment schedule ($\textbf{Z}$), the received exposure ($A_k$), the outcome ($Y_k$), the individual-specific random variable ($U$) and the time-varying covariate ($L_k$) in a scenario where the time-varying covariate induces a time trend and no carryover exists, for $t>2$. Nodes represent variables, directed edges represent causal relationships. Dashed nodes represent unobserved variables and green edges represent deterministic causal relationships. To avoid clutter, we do not depict explicitly the exogenous noise random variables $\varepsilon_k$ and $\gamma_k$ that affect respectively $Y_k$ and $L_k$.}
    \label{fig:dgm4}
\end{figure}

Consider another causal model. Suppose that the treatment is assigned according to \eqref{eq:A1} and the treatment schedule satisfies \eqref{eq:Z1}. The time-varying covariate $L_k$, instead, is generated according to
\begin{equation} \label{eq:L2}
  L_{k} = f^L(L_{k-1}, U_L, \gamma_{k})
\end{equation}
while the outcome is generated according to
\begin{equation} \label{eq:Y3}
Y_{k} = f^Y(L_{k}, A_{k}, L_{k-1}, U_Y, \varepsilon_{k}),
\end{equation}
where all symbols have the same meaning and interpretation as in Section \ref{sec:relaxed_scenario}, and $U$ is again the union of $U_L$ and $U_Y$.
The causal model described by \eqref{eq:A1}, \eqref{eq:L2}, and \eqref{eq:Y3} is depicted in the DAG in Figure \ref{fig:dgm4}. We will again assume the positivity condition \eqref{eq:positivity}.

It is immediately clear that this causal model is a special case of the one discussed in Section \ref{sec:relaxed_scenario}. Indeed, the only differences between the two data generating processes are that $A_k$ and $Y_{k-1}$ are omitted from the structural equation \eqref{eq:L1}, and $Y_{k-1}$ and $A_{k-1}$ are omitted from the structural equation \eqref{eq:Y2}. 

The variable $L_k$ represents now a cause of the outcome, which is not affected by the treatment (either directly or indirectly), and whose distribution changes over time. Moreover, as before, $L_k$ is a common cause of a proper subset of outcome measurements. As a result, strong stationarity \eqref{eq:stationarity1} is not guaranteed to hold. However, there are no direct or indirect carryover effects of the treatment, therefore condition \eqref{eq:carryov1} holds.
While this causal structure is a special case of the scenario described in Section \ref{sec:relaxed_scenario}, it can also be interpreted as an instance of what is commonly referred to as a “time trend". However, the use of this term is ambiguous in the literature. 
In the elaboration of the SPENT 2019 checklist for N-of-1 trials, a time trend is defined as “the change that would have occurred over time regardless of intervention” \cite{Porcino2020}. A more nuanced definition is the one from Schmid and Yang \cite{Schmid2022-oh}, who use the term time trend to indicate that “underlying forces may lead to changes in the outcome independent of treatment effects”. 
A similar idea is described in the user guide for design and implementation of N-of-1 trials for the Agency for Healthcare Research and Quality of the U.S. Department of Health and Human Services, prepared by the Brigham and Women’s Hospital DEcIDE Methods Center \cite{DEcIDE}. In this guide a (linear) secular trend is considered present if there is a “confounder whose effect on the outcome is linear with time” \cite{DEcIDE}.
While these definitions are not formal about the causal process they envision, they seem to point in the direction of a violation of a form of stationarity due to something other than a carryover effect; that is, they point to a violation due to the fact that the outcome measurements are affected by a covariate, which is not influenced by the treatment, and whose distribution changes over time. The causal model described by \eqref{eq:A1}, \eqref{eq:Z1}, \eqref{eq:L2} and \eqref{eq:Y3}, as depicted in Figure \ref{fig:dgm4}, represents a scenario in which a time trend is present.
Being a special case of the causal model in Section \ref{sec:relaxed_scenario}, identification and estimation results shown in this section are still valid. It is possible to identify and estimate the $U\text{-}CATE_k(u_i)$ using data from the individual $i$, under the same conditions. It is, however, sufficient to consider fewer variables when modeling $g^Y$ and $g^L$, because in this special case some of the variables in the conditioning sets are known to be conditionally independent.

\subsection{Adjustment for time trends}\label{sec:time_as_proxy}
Investigators often believe that there is a time trend in the outcome distribution. For example, in the sense used by Schmid and Yang, according to whom ``underlying forces may lead to changes in the outcome''. In Section \ref{sec:time_trend} we have denoted these underlying forces by the time-varying covariates $\bar{L}$. When these covariates are not measured, researchers nonetheless often seek to adjust for them. We consider two ways in which this is done. First, it is common to include smooth or low dimensional functions of time in regression models.  These functions could be  indicators of season or day of the week. Other linear terms, sinusoidal functions, or spline bases are also sometimes used \cite{Schmid2022-oh}. Second, and less commonly, researchers sometimes attempt to directly model the unmeasured $\bar L$ via a latent state model such as a Hidden Markov Model \cite{valeri2025}. 

In Appendix H we give sufficient conditions under which these approaches work for a class of data generating processes that include the one presented in Section \ref{sec:time_trend}. Specifically, we focus on the setting where the only non-treatment variables in the structural equation of the outcome are $U$ and the unobserved time-varying covariates, and the time-varying covariates are not affected by treatment.  We show that, under certain assumptions, the $U\text{-}CATE_k(u_i)$ is identified by a g-formula where we condition on a function of time, rather than on the unmeasured time-varying covariate. Finally, we discuss identification when latent-state models modeling directly the unmeasured time-varying variables are used.

\section{Randomization}
\label{sec:randomization}
Randomization of the treatment schedule is not necessary for the identification and estimation results that we presented in Sections \ref{sec:inference_basic_scenario} and \ref{sec:inference_relaxed_model}. 
Our inference results were conditioned on $\textbf{Z}$, and thus the results would also hold in settings where $\textbf{Z}$ is fixed to a constant $\textbf{z}$. For example, when the treatment schedule is predetermined and imposed on the participant, without being randomized. Also, our results are valid even if $U$ affects the assigned treatment schedule, which means that $\textbf{Z}$ is not an exogenous variable; our inference is, by design, conditioned on $U=u_i, \textbf{Z}=\textbf{z}_i$. This argument agrees with the intuition that an N-of-1 trial is a self-controlled design \cite{Hogben1953-fe, Vohra2015}, where confounding by $U$ is not a concern. 

Consider again the causal model described in Section \ref{sec:relaxed_scenario}. Suppose that treatment is assigned at every time point based on the value of $U$ and known variables e.g., $A_{k-1}$, $L_{k-1}$ and $Y_{k-1}$. Then the counterfactual quantity $\mathbb{E}(Y^{\bar{a}_k=\bar{x}_k}_k \mid U=u)$ is identified by the same g-formula described in \eqref{eq:theta} \cite{Robins1999}. This setting is consistent with an observational study where the time-varying confounders are known \cite{Robins1999}, or with an adaptive N-of-1 trial \cite{Essentials_ch16}. Because we are interested in the $U\text{-}CATE_k(u_i)$, our inference is conditional on the context $U=u_i,\textbf{Z}=\textbf{z}_i$, and in our causal models we always assumed some form of stationarity conditions, randomization is not necessary. 

However, randomization can be useful when an investigator is interested in different types of estimands or relies on different inferential strategies. Suppose, for example, that the investigator is interested in inference marginal over the possible values of $\textbf{Z}$. Without assuming stationarity, under the causal model represented in Figure \ref{fig:dgm1} and for specific N-of-1 trial designs, $\mathbb{E} (\hat{\tau_i} \mid U=u_i, \bar{\varepsilon}=\bar{\varepsilon}_i) = \frac{1}{t} \sum^{t}_{k=1} ICE_{i,k}$, that is $\hat{\tau}_i$ is an unbiased estimator of $\frac{1}{t} \sum^{t}_{k=1} ICE_{i,k}$ marginally over $\textbf{Z}$ (see Appendix E for more details).

In the N-of-1 trials literature, randomization has been described as a way to eliminate confounding due to autocorrelation and carryover effects \cite{daza2019, daza2022}, as a way to balance time-varying confounding factors \cite{Essentials_ch7, Essentials_ch16}, or as a way to minimize confounding and selection bias \cite{Essentials_ch7}. We believe that confusion often arises about the role of randomization in N-of-1 trials, due to different choices and beliefs regarding the units of inference, the estimand of interest, and the type of inference. For example, some authors seem to consider time units as the units of interest in N-of-1 trials, rather than individuals. Daza and colleagues considered the superset of periods experienced by one person as the target population of their inference \cite{daza2022}. Coherently, they defined as the estimand of interest the “average period treatment effect”, the contrast between the expected value of the outcome under treatment and no treatment in this superpopulation of time points \cite{daza2022}. This conceptualization justifies the idea that N-of-1 trials are RCTs in which time points are randomized. When time points are conceptualized as units drawn from a superpopulation of times, it feels natural to conduct inference marginal over the possible treatment schedules. This conceptualization also explains the use of the term confounding to indicate a lack of exchangeability (as defined in \cite{hernan2019}) between the time points (e.g., due to a time trend) \cite{DEcIDE}. Randomization of treatment at each time is then seen as a way to ensure exchangeability between time points \cite{daza2022}. However, this conceptualization hinges on the idea that observed time measurements can be considered i.i.d. draws from a superpopulation of near-infinite times.
Under this view, effects between time points, like carryover of treatment, can be considered “interference” between statistical units \cite{hernan2019}. For example Daza and colleagues in a superpopulation inference framework \cite{daza2022} or Liang and Recht in a randomization inference framework \cite{Liang2023} explicitly consider carryover effects as the presence of interference. Interestingly, in order to estimate causal effects from a single realization of social network data where interference between individuals is present, Ogburn et al.\ \cite{Ogburn2022} developed an estimation strategy similar to the one we propose for N-of-1 trials.

In this work, we considered individuals as the units of interest, aligning with the superpopulation inference framework commonly adopted in epidemiology \cite{hernan2019}. In line with the epidemiologic literature, we also chose the $U\text{-}CATE_k(u_i)$ as the parameter of interest, as it represents the conditional version of conventional population-level estimands \cite{Robins1999}. Furthermore, when individuals are the units of interest, an N-of-1 trial with randomized treatment schedule can be considered as a “mixed" experiment \cite{Robins2000}. In the causal models represented in Figures \ref{fig:dgm1}, \ref{fig:dgm3}, or \ref{fig:dgm4}, the variable $\textbf{Z}$ determines which experiment the participant does. Being known at the start of the trial and not affecting the parameter of interest $U\text{-}CATE_k(u_i)$, the treatment schedule $\textbf{Z}$ represents an ancillary statistic \cite{Robins2000}. Therefore, according to the conditionality principle \cite{hernan2019, Robins2000}, our statistical inference was conditional on $\textbf{Z}$.

\section{Data application}
\label{sec:data_application}

\subsection{Data description} \label{sec:data_description}

In this section, we rely on the results illustrated in the previous sections to analyze data from two N-of-1 trials. Specifically, we used data from N-of-1 trials on acne severity described in Fu et al.\ \cite{Fu2023}, which are publicly available \cite{GitHub_data}. In the original study, these data have been used, together with pictures taken during the experiment, to train convolutional neural networks for automated analysis of multimodal N-of-1 trials \cite{Fu2023}. Here, we only use the available tabular data and re-analyze two of the N-of-1 trials using the methods described in the previous sections.

\begin{figure}
\includegraphics[width=\textwidth]{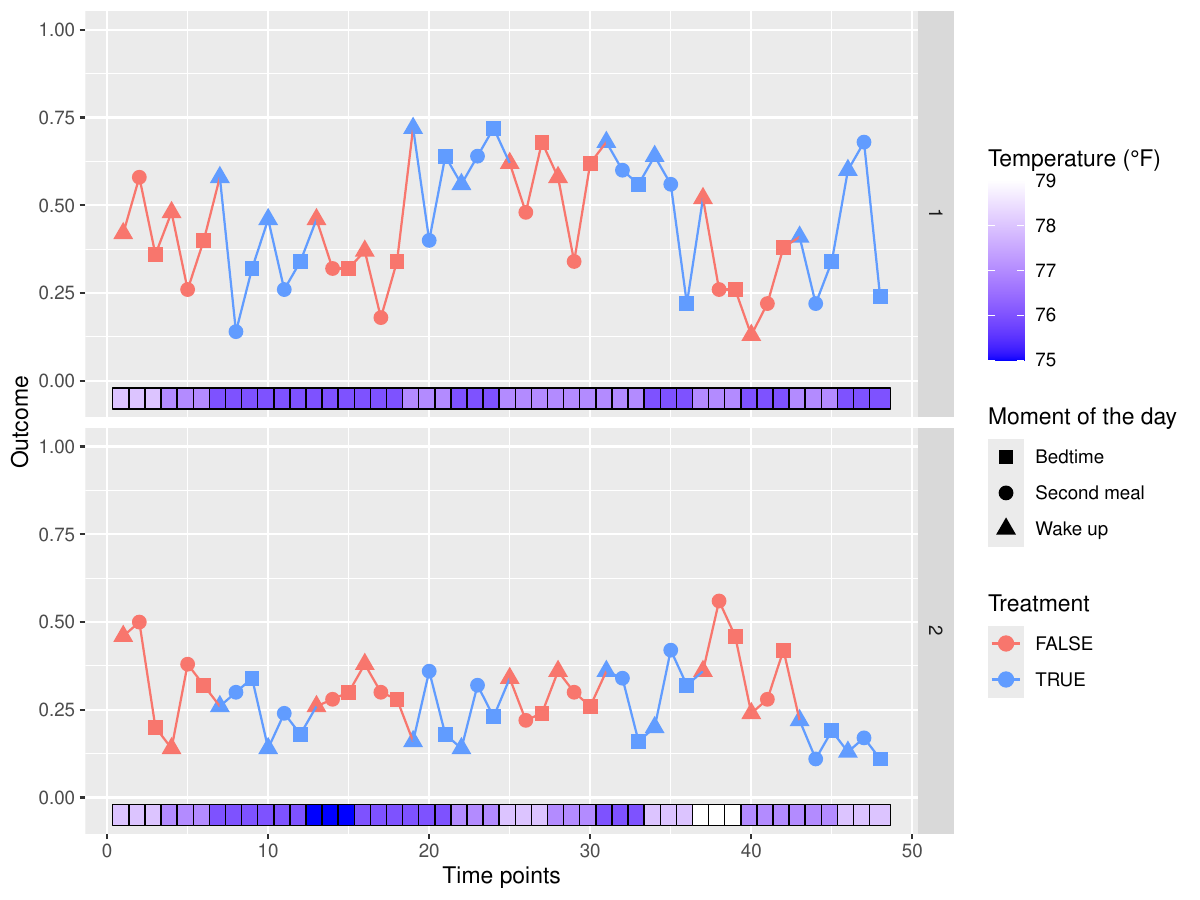}
\caption{Data from two participants (1 and 2) in the series of N-of-1 trials from Fu et al.\ \cite{Fu2023}. The two participants tested a salicylic acid-based acne control gel according to an assigned fixed treatment schedule. Acne severity is the outcome variable shown on the y-axis (ranging from 0, no visible acne, to 1, skin covered in acne). Moment of the day and daily environmental temperature are also shown.} \label{fig:4}
\end{figure}

In the series of experiments described by Fu et al.\ \cite{Fu2023}, two participants tested one acne product and three other participants tested another acne product. Here we deliberately use data from two participants (participants 1 and 2 from the original study), who tested the same product, a salicylic acid-based acne control gel \cite{Fu2023}. Both participants participated in the study for 16 days, and collected data 3 times per day: after waking up, after the second meal of the day, and before bedtime \cite{Fu2023}. Therefore, each participant $i=1,2$ contributed $t=48$ time points. These time points were subdivided into 4-day treatment cycles ($q=12$). Both participants were assigned a fixed treatment schedule $\textbf{z}=(0,0,0,0,0,0,1,1,1,1,1,1)$ which repeated itself over time. That is, for every cycle, the participant spent two days (6 time points) without treatment ($A_k=0$), and two days (6 time points) with treatment ($A_k=1$). Treatment consisted in applying the salicylic acid-based acne control gel according to a specified schedule on spots of facial or body acne \cite{Fu2023}. At every time point, following a strict protocol, a picture of the spot of facial or body acne was taken by the participants \cite{Fu2023}. The pictures were then all randomly shuffled and rated independently by five raters on a scale from 0 (no visible acne) to 1 (skin fully covered in acne) \cite{Fu2023}. Our outcome was the average of the ratings from the five raters ($Y_k$), as a measure of acne severity. We further used two additional variables ($L_k$), among the ones recorded in the original study, which might be associated with the outcome: the moment of the day (waking up, second meal, bedtime) and the recorded temperature of the environment during the day (in Fahrenheit, same value for measurements from the same study day) \cite{Fu2023}. Data about the two participants included in this analysis are visualized in Figure \ref{fig:4}. 

\subsection{Analysis}
\label{sec:analysis}

In this section, we will analyze the N-of-1 trial data described in Section \ref{sec:data_description} to assess the effect of the acne gel. For didactic purposes we will analyze the data in two different ways, imagining that in each investigation we are willing to make different assumptions about the data generating process. 

In the first analysis, we assumed the data to be generated from a simple data generating process. Specifically, we assumed the “basic scenario" causal model as described in Section \ref{sec:basic_scenario}, with no carryover, no time-varying common causes of the outcome, no outcome-outcome effects, and no time trends. Under this causal model, the outcome distribution does not change across time when considering time points with the same treatment from one individual. To test if this data generating process is plausible, we fit a beta regression model \cite{Ferrari2004,CribariNeto2010} with acne severity as the dependent variable and time point number as the independent variable, separately among treated and non-treated time points. We did not reject the null hypothesis of null coefficients for the time variable in either of the two regression models at a 5\% Type I error rate, both for participant 1 (p=0.734 and p=0.399) and participant 2 (p=0.159 and p=0.405). Therefore, we did not reject the hypothesis that the data were generated from the basic causal model.

Under the basic causal model, we know that $\hat{\tau}_1$ and $\hat{\tau}_2$, as defined in \eqref{eq:tauhat1}, are valid estimators of $U\text{-}CATE_k(u_1)$ and $U\text{-}CATE_k(u_2)$, which are constant over time. Using the result in \eqref{eq:basic_z_distribution}, we estimated the effect of treatment for participant 1 to be $\hat{\tau}_1$=0.081 (95\%CI: -0.013 -- 0.175). The estimated effect of treatment for participant 2 had the opposite sign, $\hat{\tau}_2$=\,-0.094 (95\%CI: -0.148 -- -0.040), favoring the treatment. In this first investigation, assuming the basic causal model, we would therefore reject the null hypothesis of no individual-specific effect for participant 2, while we would not reject the null hypothesis for participant 1.

\begin{figure}
\centering
\includegraphics[height=9cm]{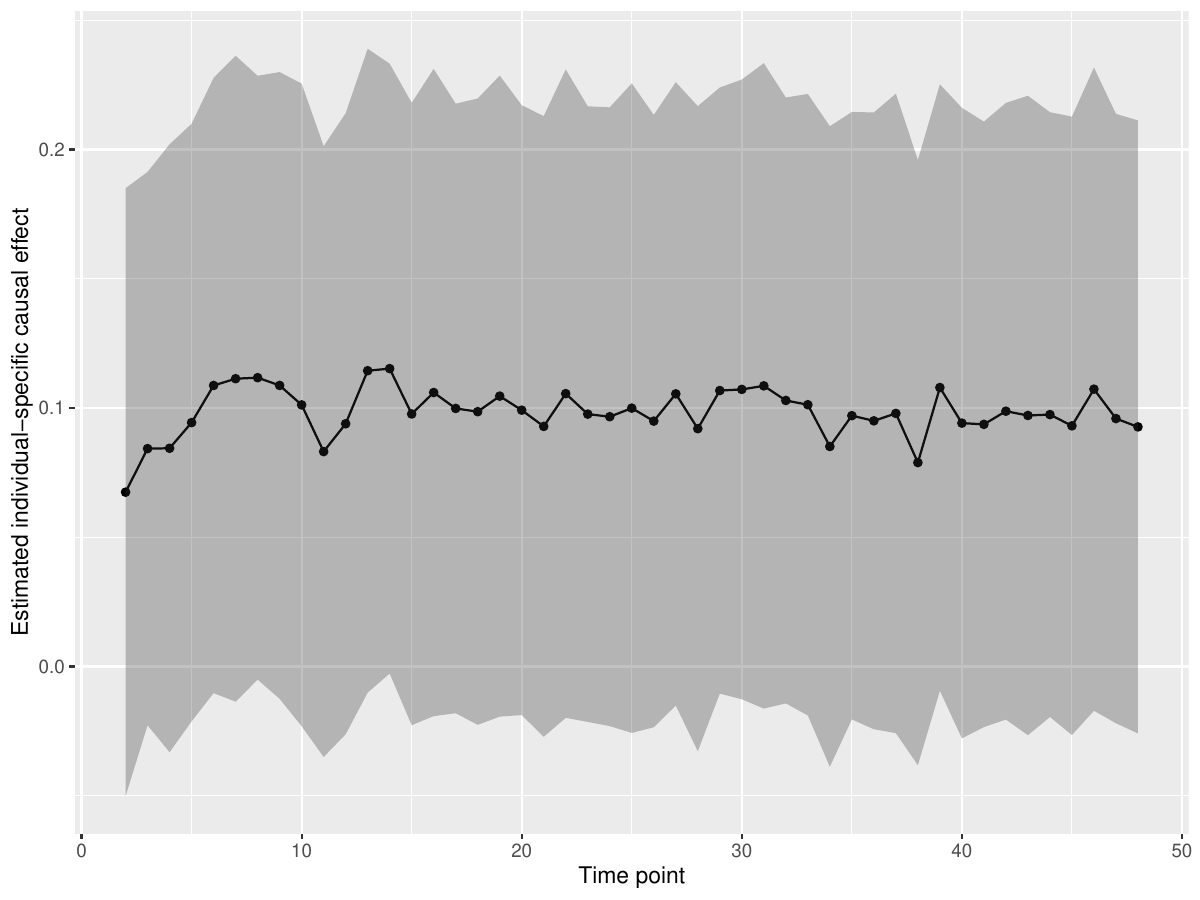}
\qquad
\centering
\includegraphics[height=9cm]{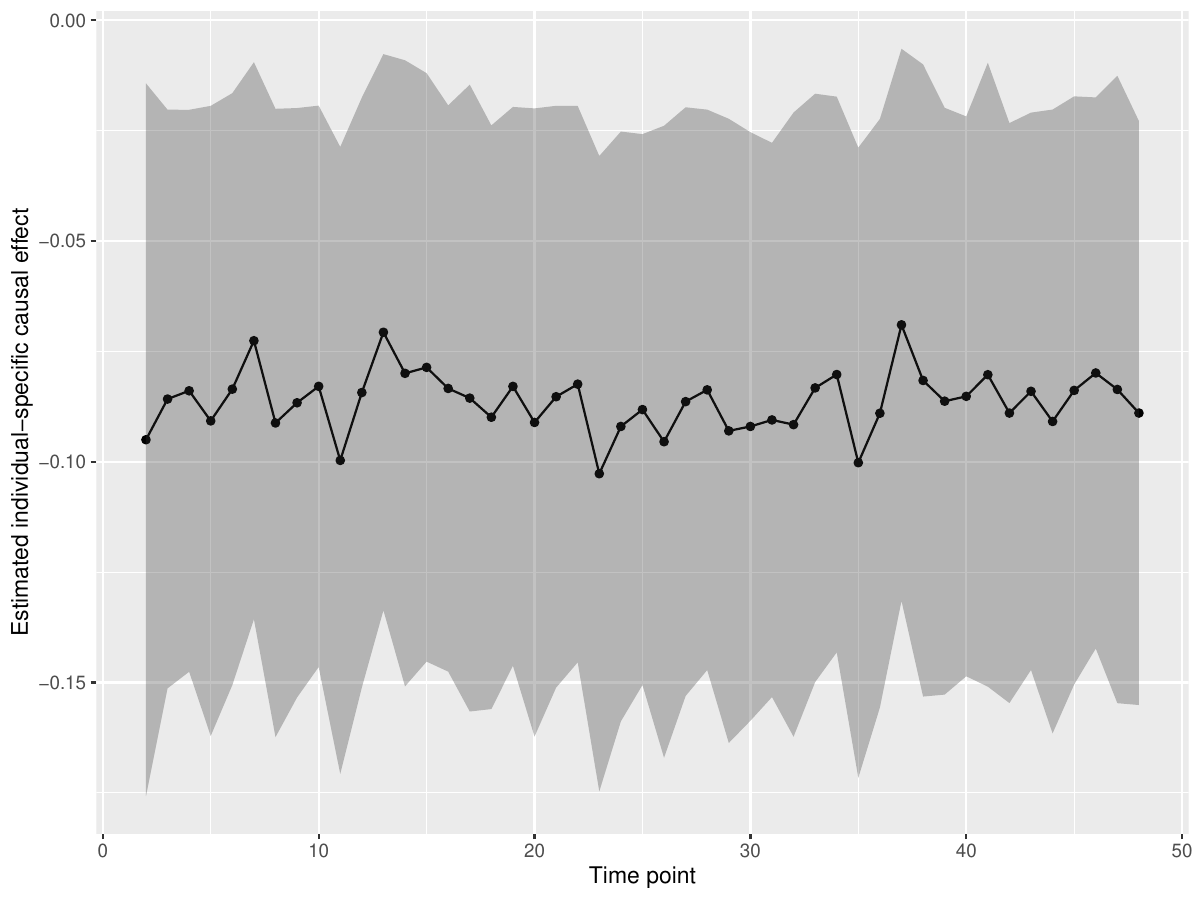}
\caption{Results from the second data analysis. Estimates of the individual-specific causal effects for the first participant (upper panel) and for the second participant (lower panel) for all time points as obtained from the g-computation algorithm.}
\label{fig:5}
\end{figure}

In a second analysis, we relaxed the causal assumptions. We assumed that time of day and temperature could be associated with acne severity. We could rule out that the treatment status or the outcome affect these two time-varying covariates. Moreover, current acne status might affect acne manifestation in the future, and the treatment effect might carry over to future times, but we assumed that the direct carryover and outcome-outcome effects do not last longer than a few hours. This motivates a special case of the causal model from Section \ref{sec:relaxed_scenario}. 
In this setting, we have direct carryover of the treatment, outcome-outcome effect, and the presence of covariates (temperature and time of day) that influence themselves and the outcome. All these relationships can span up to one time point in the future. To estimate the individual-specific causal effects for both participants in this complex scenario, we relied on a special case of the g-formula presented in Equation \eqref{eq:theta}. To estimate $\hat{\theta}^{x}_{i,k}$, we fit two regression models, a beta regression for the conditional distribution of the outcome and an ordinal logistic regression for the conditional distribution of the temperature. In the beta regression, the outcome at time $k$ was the dependent variable while treatment at time $k$, treatment at time $k-1$, temperature of the day for time $k$, moment of the day at $k$, and outcome at time $k-1$ were the independent variables. Since temperature was a daily measurement, the ordinal logistic regression was only fit for the first measurement of the day and to avoid overfitting we used a penalized
maximum likelihood estimation approach \cite{harrell2001regression} with an a priori small penalty of 0.5. In the regression, temperature was the dependent variable, while temperature in the previous day was the independent variable.
In order to always have a well-defined instantiation of all variables at time $k-1$, we took the values of the variables at the first time point to be fixed (see comments in Section \ref{sec:relaxed_scenario}). Therefore, we implicitly conditioned on the information obtained at the first time point, and focused on estimating the individual-specific effect from time 2 to the end of the experiment. No regression model was required for the other covariate, moment of the day, as this variable is deterministically obtained based on its previous value, and is therefore fixed conditioning on the first time point. Instead of analytically solving the g-formula in Equation \eqref{eq:theta} for each participant and then estimating the individual-specific causal effects, we relied on a g-computation Monte Carlo simulation algorithm. For each participant, we generated a dataset under the “always treated” strategy and a dataset under the “never treated” strategy using the beta and ordinal logistic regression models to generate the data at every time sequentially. We then calculated the differences between the outcome in both scenarios for each time point. This procedure was repeated 500 times, and the average of the differences across the 500 repetitions for the same time $k$ was considered an approximation of $\hat{\theta}^{1}_{i,k}-\hat{\theta}^{0}_{i,k}$ and, therefore, a point estimate of $U\text{-}CATE_k(u_i)$. The pointwise 95 \% confidence intervals were obtained by parametric bootstrap. 500 samples were generated using the beta and ordinal logistic regression models fitted in the original sample. These bootstrapped samples were obtained fixing the treatment schedule to be the one of the observed study participants. In each bootstrapped sample, the procedure to obtain the individual-specific causal effect point estimate was repeated. The standard deviation of the estimates for each time point across the repetitions was considered to be an approximation of the standard error of the estimator and used to obtain confidence intervals with normal approximation. In Figure \ref{fig:5} we reported the estimated individual-specific causal effects for both participants at all times. Point estimates of the individual-specific causal effect ranged over time between 0.067 and 0.115 for participant 1, and between -0.103 and -0.069 for participant 2. 

In this investigation, we found a favorable effect of the “always treat” strategy compared to the “never treat” strategy for participant 2 for all time points, while no favorable effect was observed for participant 1.

Interestingly, both investigations concluded that the salicylic acid-based acne control gel seems to have a favorable effect on acne severity for participant 2, while no evidence of an effect for the same treatment was found for participant 1. This simple example shows the importance of N-of-1 trials when effect heterogeneity exists.

The R code for all analyses can be found at \url{https://github.com/MarcoPic-dot/Causal_Inference_Nof1_trial_acne}. This code can be readily adapted or extended to reflect different causal models, time intervals, or parametric assumptions.

\section{Relevance of N-of-1 trial results} \label{sec:relevance_results}

We have considered inference from an N-of-1 trial as conditional on $U$. Since $U$ is likely a high dimensional variable that represents the characteristics of the individual under study, it is plausible that causal effects estimated in an N-of-1 trial are only relevant for the individual under study, and have limited generalizability to other individuals. This is a recognized characteristic of N-of-1 trials \cite{Guyatt1986-je, Mirza2017-uh, Lillie2011-gv}, and the personalized nature of this study design is considered its strength \cite{Schork2018-xs, Lillie2011-gv, Guyatt1986-je, Essentials_ch6, Schmid2022-oh}. In a clinical setting, an N-of-1 trial is often expected to result in a decision: should the patient, from the end of the trial onwards, take the treatment or the control? 

Consider, for example, the case study reported by Guyatt et al.\ \cite{Guyatt1986-je} in the article that popularized the use of N-of-1 trials. They reported an N-of-1 trial carried out in a clinical setting with the aim of determining if a patient should in the future use certain drugs or not to treat his asthma \cite{Guyatt1986-je}. The interest of the clinician and the patient in this trial was in determining if always taking the treatment or the control from a moment onwards results in better outcomes. This type of implicit causal contrast also motivates our choice of a causal estimand, the $U\text{-}CATE_k(u)$, that compares an “always treatment" strategy against an “always control" strategy. Estimating the $U\text{-}CATE_k(u_i)$ at all time points clarifies whether the patient with $U=u_i$ would have benefited from taking treatment for the whole duration of the study. However, this is not necessarily the answer to the clinical question of interest, which concerns effects \textit{after} the trial is conducted. Indeed, the validity of the $U\text{-}CATE_k(u_i)$ estimate for guiding treatment decisions after the study ends relies on additional assumptions.

Imagine that the true data generating process is the one described in Section \ref{sec:basic_scenario}, and we conduct an N-of-1 trial that provides us with a precise estimate of the $U\text{-}CATE_k(u_i)$. This estimate is not guaranteed to represent the causal contrast between the always-treat and the never-treat strategies for times $k>t$. Under the strong assumption that condition \eqref{eq:stationarity1_expectation} continues to hold beyond time $t$, the estimated individual-specific causal effect would be a valid estimate also for e.g., $U\text{-}CATE_{t+1}(u_i)$. The plausibility of this stationarity assumption, however, depends on the time span.

In more complex data generating processes, such as the ones described in Sections \ref{sec:relaxed_scenario} and \ref{sec:time_trend}, transportability of the individual-specific causal effect to the future is more challenging. In particular, the $U\text{-}CATE_{k}(u_i)$ is not constant over time in this setting. Different initial conditions in the time-varying variables may lead to a different value of the individual-specific causal effects over time for a period of length $t$ even assuming that stationarity conditions \eqref{eq:stationarityL} and \eqref{eq:stationarity2} hold. If these stationarity assumptions continue to hold, however, it is possible to use the g-formula in \eqref{eq:theta} to predict what would happen after the trial.

We emphasize that assessing the relevance (generalizability) of experiment results in the future is also a challenge in traditional RCTs, and is not unique to N-of-1 trials. Learning from the past to act in the future requires implicitly some notion of stationarity. The N-of-1 trial design, however, has some specific attributes that may be exploited to test some assumptions. If, after the trial phase, the patient receives only one version of the treatment, the data collected in this post-trial phase may be used to test the stationarity conditions (under one version of the treatment) and whether the counterfactual predictions were valid.

\section{Aggregating results from series of N-of-1 trials}
\label{sec:aggregating_series}

Under explicit assumptions and data structures, N-of-1 trials can also be used to estimate population-level causal effects, such as the $ACE_k$ defined in \eqref{eq:ACE} \cite{Zucker1997, Essentials_ch4}.

To illustrate a point, suppose first that there is no effect modification by $U$ on the additive scale, and thus
\begin{equation} \label{eq:homogeneity}
\mathbb{E}(Y^{\bar{a}_k=\bar{1}_k}_{k} \mid U=u) - \mathbb{E}(Y^{\bar{a}_k=\bar{0}_k}_{k} \mid U=u) = \mathbb{E}(Y^{\bar{a}_k=\bar{1}_k}_{k}) - \mathbb{E}(Y^{\bar{a}_k=\bar{0}_k}_{k}) 
\end{equation}
for every $k$ and every $u \in \mathcal{U}$. When the equality in \eqref{eq:homogeneity} holds, $U\text{-}CATE_k(u)$ and $ACE_k$ coincide, and the estimate of $U\text{-}CATE_k(u_i)$ obtained from a single N-of-1 trial of an individual $i$ can be used to estimate the average causal effect in the population. Indeed, if we know that the effect of the treatment is the same regardless of the characteristics $U$, identifying an effect for a specific set of characteristics is sufficient to identify the causal effect at the population level. 
Of course, assumption \eqref{eq:homogeneity} is not realistic, and effect heterogeneity is exactly what historically motivated the use of N-of-1 trials \cite{Tabery2011, Hogben1953-fe, Schmid2022-oh, Essentials_ch4, Essentials_ch6, Essentials_ch11}. 

Yet, when Equation \eqref{eq:homogeneity} fails, we can aggregate a series of N-of-1 trials to estimate $ACE_k$. Consider the basic causal model described in Section \ref{sec:basic_scenario}. Under this causal model, $ACE_k$ is constant over time, and we show in Appendix F that the average of $\hat{\tau_i}$ across a sample of $n$ individuals from the population of interest, $\frac{1}{n} \sum^n_{i=1} \hat{\tau_i}$, is an unbiased and consistent estimator (as $n$ goes to infinity) for $ACE_k$. Similarly, under the relaxed causal model described in Section \ref{sec:relaxed_scenario}, the average of the individual-specific effect estimates across $n$ individuals,
\begin{equation*}
\frac{1}{n} \sum^{n}_{i=1}(\hat{\theta}^{1}_{i,k}-\hat{\theta}^{0}_{i,k}),
\end{equation*}
converges towards the average causal effect $ACE_k$ when sequentially $t$ and $n$ go to infinity (see Appendix G). These results imply that, under the assumed causal models, we can estimate population-level effects, equal to those targeted in conventional parallel group RCTs, by aggregating large series of N-of-1 trials.

When interest lies in population-level effects, such as $ACE_k$, stationarity assumptions are not necessary for valid inference when treatment is randomized. Consider a data generating process compatible with a parallel-group RCT where we randomize individuals to two possible treatment schedules, $\textbf{z}=\bar{1}_q$ and $\textbf{z}=\bar{0}_q$. Specifically consider the causal model in Figure \ref{fig:dgm3} under a FFRCISTG interpretation, and treatment assignment given in Equation \eqref{eq:A1}. Randomization of the treatment schedule across individuals ensures exchangeability between individuals who are assigned to a treatment schedule and individuals who are assigned to the other treatment schedule \cite{hernan2019}. The mean outcome of a group of individuals who happened, by randomization, to receive treatment at all times until $k$ is a valid estimator of the average counterfactual outcome at $k$ under the always treatment strategy, for every $k$. Therefore, this design allows for estimation of the $ACE_k$ at any point in time \cite{hernan2019}. 

Consider, instead, a data generating process compatible with a series of N-of-1 trials or equivalently with a crossover randomized trial. Imagine that no carryover effect of the treatment exists (i.e., condition \eqref{eq:carryov1} holds). For example, consider the causal model in Figure \ref{fig:dgm4} under a FFRCISTG interpretation, and treatment assignment given in Equation \eqref{eq:A1}. Imagine that we randomize individuals to $\mathcal{Z} = \{(1,0,1),(0,1,0)\}$. Here, we do not require stationarity. Because we have exchangeability at every time point, we can consider this design as a series of $t$ RCTs that assign treatment at a single time point. The mean difference between the outcomes of the treated and the non-treated individuals at a given time $k$ is an unbiased and consistent estimator of $ACE_k$. We do not need to measure the time-varying covariate $L_k$ in this setting. This simple example illustrates that, assuming no carryover effects, we can use data from a series of N-of-1 trials with a randomized treatment schedule to estimate population-level effects, without assuming any form of stationarity.

\section{Conclusion} 
\label{sec:conclusion}

In this work we grounded N-of-1 trials in a formal causal inference framework. Building on previous work by Robins et al.\ \cite{Robins1999}, we committed to an explicit statistical and causal framework to motivate the analysis of N-of-1 trials. We defined the $U\text{-}CATE_k(u)$, a conditional causal effect that represents a causal estimand for N-of-1 trials. We discussed identifiability and estimation of this individual-specific causal effect in a basic scenario characterized by a strong stationarity assumption, and complex scenarios where carryover effects, time trends, time-varying common causes of the outcome, and outcome-outcome effects are allowed. We also briefly discussed the role of randomization, the clinical relevance of N-of-1 trial results, and the aggregation of series of N-of-1 trials for inference about population-level effects typically targeted in RCTs.

N-of-1 trials are often motivated by settings where there is therapeutic uncertainty, the treatment has highly heterogeneous effects, the condition is chronic and stable, the treatment has short term consequences with rapid onset of effects, the treatment does not have carryover effects (no long term consequences), and the relevant outcome can be easily measured \cite{Schmid2022-oh, Essentials_ch2, Essentials_ch4, Essentials_ch16, Mirza2017-uh}. We have shown that when no carryover effects, no trends, no outcome-outcome effects, and no time-varying common causes of the outcome are present, simple analyses of N-of-1 trials might be justified. G-methods \cite{Robins1999} may suffice under more complex scenarios. It is, therefore, important when analyzing N-of-1 trials to explicitly represent the assumptions behind the data generating process. 

We have committed to a frequentist statistical framework based on superpopulation inference, like most statistical inference conducted in epidemiology and medical research. Our choice is mathematically consistent, but requires commitment to certain subtle ideas. We make inference about a hypothetical superpopulation of individuals with the same individual-specific variable $U=u_i$ as the trial participant, which in reality typically does not exist. Randomization-based inference frameworks \cite{Bojinov2019, Malenica2023, Liang2023} might also be used to formalize and analyze N-of-1 trials, without the need of envisioning a superpopulation. However, unlike the randomization-based frameworks, the superpopulation framework allows us to explicitly evaluate transportability to different time periods and to make precise connections between effects at the individual, subpopulation and the population level. Moreover, in some N-of-1 trials, such as the one analyzed in Section \ref{sec:data_application}, the treatment schedule is arbitrarily fixed a priori, and is not randomly assigned. In these situations, the interpretation of randomization-based inferences is ambiguous \cite{Zhang2023}. We have instead shown that, in our framework, inference is also possible when the treatment schedule is fixed to a specific value; this is because our inference is conditional on the treatment schedule and we rely on stationarity assumptions.

\subsection{Future work}
\label{sec:future_work}

In this work, we only considered certain specific data generating processes. Several other causal models are possible. For example, we considered a specific treatment assignment function where treatment status is assigned repeatedly over time, according to a fixed treatment schedule. Our treatment assignment mechanism covers all possible treatment sequences for a finite time interval, and therefore is compatible with other designs of randomized N-of-1 trials in finite time. However, it does not cover all possible treatment sequences that can be generated over infinite time horizons, such as treatments generated by Bernoulli trials at any time point \cite{Bojinov2019} and paired cycle designs \cite{Senn2024}. Furthermore, we only briefly considered designs compatible with adaptive N-of-1 trials where treatments are assigned over time as a function of previous treatments, covariates and outcomes. Moreover, we have assumed full compliance and ignored issues related to loss to follow-up or missing data. With minor extensions, our work can accommodate these complications.

A possible direction for future research is the application to individualized therapies \cite{Kim2019}. Recently, the U.S. Food and Drug Administration released a draft of a regulatory framework for individualized therapies in rare genetic diseases \cite{Gottlieb2026}.  Collecting evidence for the effectiveness of these so-called ``N-of-1 therapies'' is challenging \cite{Gottlieb2026}. These therapies are engineered and tailored to address unique genetic perturbations that affect only a few patients, or even just one patient. Thus, they cannot be feasibly tested in conventional RCTs \cite{Gottlieb2026}, but N-of-1 trials offer a potential solution because they require only a single individual \cite{KimMcManus2024}. However, crossover designs, including N-of-1 trials, must also be carefully conducted and interpreted, because these therapies can have long-lasting effects and are sometimes administered in a single dose \cite{KimMcManus2024}. Conceptually, once the therapy is initiated, we cannot observe a representative value of the outcome variable under a non-therapy state again, due to long-lasting effects. Therefore, inference about the effect of the therapy might be limited to the comparison of outcomes before and after therapy initiation. In finite time, our work covers interventional pre-post studies \cite{Aggarwal2019} as a special case, where one individual is first observed without therapy, and then observed after therapy initiation, following a pre-specified schedule. Therefore, our results could be useful in assessing N-of-1 therapies. As a simple example, consider a single dose therapy, and suppose that the investigators collect information over time about an outcome. The time of therapy administration is chosen in advance and there is no confounding. Under the causal model and stationarity assumption described in Section \ref{sec:basic_scenario}, our results can be used to identify the individual-specific effect. The approach described in sections \ref{sec:complex_scenarios} and \ref{sec:inference_relaxed_model} can also be adapted to estimate relevant individual-specific effects for single dose therapy when time-varying covariates affect the outcome.

The use of formal causal arguments in the analysis of medical N-of-1 trials is scarce \cite{daza2019, daza2022}. Most N-of-1 trials are analyzed using traditional parametric models, such as linear mixed models \cite{Essentials_ch12}, relying on strong assumptions with an unclear causal justification. In particular, the estimands, and sufficient identifiability conditions, are often implicit and vague. The results in this work can help clarify the causal interpretation of these existing methods.

\section*{Funding}
Mats J. Stensrud was funded by the Swiss National Science Foundation (SNSF Starting Grants, Grant number: 211550).

\section*{Acknowledgments}
The authors thank Julien D. Laurendeau for helpful comments on earlier drafts of this manuscript. During the preparation of this manuscript the authors used ChatGPT in order to assist with formatting, language editing, and error checking. The authors reviewed and edited the content as needed and take full responsibility for the content of the publication.

\bibliographystyle{unsrt.bst}
\bibliography{references}

\newpage
\parindent=0pt
\section*{Appendix A}

Consider two generic time points $k_*$ and $k$, and a value $u \in \mathcal{U}$. We have that
\begin{equation*}
\begin{split}
      U\text{-}CATE_{k_{*}}(u) &\overset{\eqref{eq:ucate1}}{=} \mathbb{E}(Y^{\bar{a}_{k_{*}} = \bar{1}_{k_{*}}}_{{k_{*}}} \mid U=u) - \mathbb{E}(Y^{\bar{a}_{k_{*}} = \bar{0}_{k_{*}}}_{{k_{*}}} \mid U=u) \\
     &\overset{\eqref{eq:carryov1}}{=} \mathbb{E}(Y^{a_{k_{*}} = 1}_{{k_{*}}} \mid U=u) - \mathbb{E}(Y^{a_{k_{*}} = 0}_{{k_{*}}} \mid U=u) \\
     &\overset{\eqref{eq:stationarity1_expectation}}{=} \mathbb{E}(Y^{a_{k} = 1}_{{k}} \mid U=u) - \mathbb{E}(Y^{a_{k} = 0}_{{k}} \mid U=u) \\
     &\overset{\eqref{eq:carryov1}}{=} \mathbb{E}(Y^{\bar{a}_k = \bar{1}_k}_{k} \mid U=u) - \mathbb{E}(Y^{\bar{a}_k = \bar{0}_k}_{k} \mid U=u) \overset{\eqref{eq:ucate1}}{=} U\text{-}CATE_k(u).
\end{split}
\end{equation*} 
Therefore the $U\text{-}CATE_k(u)$ is constant over time under the basic causal model.

Consider $k^{\textbf{z}}_1 \in \{k_* \in \{1,...,t\} \mid f^{A}(\textbf{z},k_*)=1\}$ and $k^{\textbf{z}}_0 \in \{k_* \in \{1,...,t\} \mid f^{A}(\textbf{z},k_*)=0\}$. That is $k^{\textbf{z}}_1$ and $k^{\textbf{z}}_0$ are two time points such that $A_{k^{\textbf{z}}_1}=1$ and $A_{k^{\textbf{z}}_0}=0$ when $\textbf{Z}=\textbf{z}$. Then,
\begin{equation*}
\begin{split}
     &\tau(u,\textbf{z}) \overset{\eqref{eq:tau1}}{=} \mathbb{E}(Y_{k^{\textbf{z}}_1} \mid A_{k^{\textbf{z}}_1}=1, U=u, \textbf{Z}=\textbf{z}) - \mathbb{E}(Y_{k^{\textbf{z}}_0} \mid A_{k^{\textbf{z}}_0}=0, U=u, \textbf{Z}=\textbf{z}) \\   &\overset{\eqref{eq:consistency}}{=} \mathbb{E}(Y^{a_{k^{\textbf{z}}_1} = 1}_{k^{\textbf{z}}_1} \mid A_{k^{\textbf{z}}_1}=1, U=u, \textbf{Z}=\textbf{z}) - \mathbb{E}(Y^{a_{k^{\textbf{z}}_0} = 0}_{k^{\textbf{z}}_0} \mid A_{k^{\textbf{z}}_0}=0, U=u, \textbf{Z}=\textbf{z})\\
     &\overset{(Y^{a_k}_{k} \independent A_k,\textbf{Z} \mid U)}{=} \mathbb{E}(Y^{a_{k^{\textbf{z}}_1} = 1}_{k^{\textbf{z}}_1} \mid U=u) - \mathbb{E}(Y^{a_{k^{\textbf{z}}_0} = 0}_{k^{\textbf{z}}_0} \mid U=u)\\
     &\overset{\eqref{eq:stationarity1_expectation}}{=} \mathbb{E}(Y^{a_k = 1}_{k} \mid U=u) - \mathbb{E}(Y^{a_k = 0}_{k} \mid U=u)\\
     &\overset{\eqref{eq:carryov1}}{=} \mathbb{E}(Y^{\bar{a}_k = \bar{1}_k}_{k} \mid U=u) - \mathbb{E}(Y^{\bar{a}_k = \bar{0}_k}_{k} \mid U=u) \overset{\eqref{eq:ucate1}}{=} U\text{-}CATE_k(u)
\end{split}
\end{equation*} 
In the expressions above, all quantities are well defined because condition \eqref{eq:Z1} ensures that $k^{\textbf{z}}_1$ and $k^{\textbf{z}}_0$ exist regardless of the value of $\textbf{z}$.

\section*{Appendix B}

We will use that, for $x \in \{0,1\}$,
\begin{equation*}
    \begin{split}
        &\frac{1}{\sum^{t}_{k=1} \{ \mathbb{I}(A_{i,k}=x) \}} \sum^{t}_{k=1} \{ \mathbb{I}(A_{i,k}=x) Y_{i,k} \} \\
        &\overset{\eqref{eq:A1}}{=} \frac{1}{\sum^{t}_{k=1} \{ \mathbb{I}(f^{A}(\textbf{Z}_i,k)=x) \}} \sum^{t}_{k=1} \{ \mathbb{I}(f^{A}(\textbf{Z}_i,k)=x) Y_{i,k} \}\\
        &\overset{\eqref{eq:Y1}}{=} \frac{1}{\sum^{t}_{k=1} \{ \mathbb{I}(f^{A}(\textbf{Z}_i,k)=x) \}} \sum^{t}_{k=1} \{ \mathbb{I}(f^{A}(\textbf{Z}_i,k)=x) f^Y(A_{i,k}, U_i, \varepsilon_{i,k}) \} \\
        &= \frac{1}{\sum^{t}_{k=1} \{ \mathbb{I}(f^{A}(\textbf{Z}_i,k)=x) \}} \sum^{t}_{k=1} \{ \mathbb{I}(f^{A}(\textbf{Z}_i,k)=x) f^Y(x, U_i, \varepsilon_{i,k}) \}.
    \end{split}
\end{equation*}

\subsection*{Unbiasedness}

Consider the following expectation conditioned on $U=u_i$ and $\textbf{Z}=\textbf{z}_i$,
\begin{equation*}
\begin{split}
    &\mathbb{E} \bigg( \frac{1}{\sum^{t}_{k=1} \{ \mathbb{I}(f^{A}(\textbf{Z},k)=x) \}} \sum^{t}_{k=1} \{ \mathbb{I}(f^{A}(\textbf{Z},k)=x) f^Y(x, U, \varepsilon_{k}) \} \bigg| U=u_i, \textbf{Z}=\textbf{z}_i \bigg) \\ 
    &=\frac{1}{\sum^{t}_{k=1} \{ \mathbb{I}(f^{A}(\textbf{z}_i,k)=x) \}} \sum^{t}_{k=1} \{ \mathbb{I}(f^{A}(\textbf{z}_i,k)=x) \mathbb{E} ( f^Y(x, U, \varepsilon_{k}) \mid U=u_i, \textbf{Z}=\textbf{z}_i ) \} \\
    &=\frac{1}{\sum^{t}_{k=1} \{ \mathbb{I}(f^{A}(\textbf{z}_i,k)=x) \}} \sum^{t}_{k=1} \{ \mathbb{I}(f^{A}(\textbf{z}_i,k)=x) \mathbb{E} ( Y^{a_k=x}_k \mid U=u_i, \textbf{Z}=\textbf{z}_i ) \} \\
    &\overset{(Y^{a_k}_k \independent \textbf{Z} \mid U)}{=} \frac{1}{\sum^{t}_{k=1} \{ \mathbb{I}(f^{A}(\textbf{z}_i,k)=x) \}} \sum^{t}_{k=1} \{ \mathbb{I}(f^{A}(\textbf{z}_i,k)=x) \mathbb{E} ( Y^{a_k=x}_k \mid U=u_i ) \} \\
    &\overset{\eqref{eq:stationarity1_expectation}}{=} \frac{1}{\sum^{t}_{k=1} \{ \mathbb{I}(f^{A}(\textbf{z}_i,k)=x) \}} \mathbb{E} ( Y^{a_{k_{*}}=x}_{k_*} \mid U=u_i ) \sum^{t}_{k=1} \{ \mathbb{I}(f^{A}(\textbf{z}_i,k)=x) \} \\
    &= \mathbb{E} ( Y^{a_{k_*}=x}_{k_*} \mid U=u_i )
\end{split}
\end{equation*}
with $k_*$ being a generic time point.

Therefore, 
\begin{equation*}
    \mathbb{E} (\hat{\tau_i} \mid U=u_i, \textbf{Z}=\textbf{z}_i) = \mathbb{E} ( Y^{a_{k_*}=1}_{k_*} \mid U=u_i ) - \mathbb{E} ( Y^{a_{k_*}=0}_{k_*} \mid U=u_i ) \overset{(A)}{=} \tau(u_i, \textbf{z}_i). 
\end{equation*}

\subsection*{Convergence for increasing number of time points}

Given the outcome generating process \eqref{eq:Y1}, we have that
\begin{equation*}
\frac{1}{\sum^{t}_{k=1} \{ \mathbb{I}(f^{A}(\textbf{z}_i,k)=x) \}} \sum^{t}_{k=1} \{ \mathbb{I}(f^{A}(\textbf{z}_i,k)=x) f^Y(x, u_i, \varepsilon_{k}) \}
\end{equation*}
 is an average of $\sum^{t}_{k=1} \{ \mathbb{I}(f^{A}(\textbf{z}_i,k)=x) \}$ i.i.d.\ random variables with the same mean $\mathbb{E} ( Y^{a_{k_{*}}=x}_{k_*} \mid U=u_i )$. 
 
 According to condition \eqref{eq:Z1}, for any $x$, $\sum^{t}_{k=1} \{ \mathbb{I}(f^{A}(\textbf{z}_i,k)=x) \}$ is non-decreasing in $t$ and diverges to infinity. Therefore, the weak law of large numbers ensures that
\begin{equation*}
\frac{1}{\sum^{t}_{k=1} \{ \mathbb{I}(f^{A}(\textbf{z}_i,k)=x) \}} \sum^{t}_{k=1} \{ \mathbb{I}(f^{A}(\textbf{z}_i,k)=x) f^Y(x, u_i, \varepsilon_{k}) \} \xrightarrow{P} \mathbb{E} ( Y^{a_{k_{*}}=x}_{k_*} \mid U=u_i )
\end{equation*}
when $t\to\infty$, with $k_*$ being a generic time point. This means that
\begin{equation*}
\begin{split}
\hat{\tau_i} &\xrightarrow{P} \mathbb{E} ( Y^{a_{k_{*}}=1}_{k_*} \mid U=u_i ) - \mathbb{E} ( Y^{a_{k_{*}}=0}_{k_*} \mid U=u_i ) \overset{(A)}{=} \tau(u_i, \textbf{z}_i)
\end{split}
\end{equation*}
when $t\to\infty$, for a fixed context $U=u_i, \textbf{Z}=\textbf{z}_i$.

Moreover, because the outcome observations for the same $U=u_i$ and the same level of treatment $x$ have the same variance $\sigma^2(x,u_i)$, the Central Limit Theorem ensures that
\begin{equation*}
\begin{split}
\sqrt{\sum^{t}_{k=1} \{ \mathbb{I}(f^{A}(\textbf{z}_i,k)=x) \}} \bigg( \frac{\sum^{t}_{k=1} \{ \mathbb{I}(f^{A}(\textbf{z}_i,k)=x) f^Y(x, u_i, \varepsilon_{k}) \}}{\sum^{t}_{k=1} \{ \mathbb{I}(f^{A}(\textbf{z}_i,k)=x) \}} - \mathbb{E} ( Y^{a_{k_{*}}=x}_{k_*} \mid U=u_i ) \bigg) \\
\xrightarrow{d} \mathcal{N}(0, \sigma^2(x,u_i))
\end{split}
\end{equation*}
as $t$ grows. Since $\hat{\tau_i}$ is the sum of two independent quantities that can be approximated by normal distributions, we have that
\begin{equation*}
    \hat{\tau_i} \mid U=u_i, \textbf{Z}=\textbf{z}_i \approx \mathcal{N}\bigg(\tau(u_i,\textbf{z}_i), \frac{\sigma^2(1,u_i)}{\sum^{t}_{k=1} \{ \mathbb{I}(f^{A}(\textbf{z}_i,k)=1) \}} + \frac{\sigma^2(0,u_i)}{\sum^{t}_{k=1} \{ \mathbb{I}(f^{A}(\textbf{z}_i,k)=0) \}} \bigg).
\end{equation*}

To provide an approximation of the above conditional variance, define $$\alpha = \frac{\sum^q_{k=1}z_{i,k}}{q}= \frac{\sum^q_{k=1} \{f^{A}(\textbf{z}_i,k) \}}{q}.$$ 

Consider the quantity
\begin{equation*}
\begin{split}
    &\sum^{t}_{k=1} \{ \mathbb{I}(f^{A}(\textbf{z}_i,k)=1) \} = \sum^{t}_{k=1} \{ f^{A}(\textbf{z}_i,k) \} \\
    &= 
    \begin{cases}
      \alpha \cdot t & \text{if $(t \text{ mod } q)=0$}\\
      \alpha \cdot (t - (t \text{ mod } q)) + \sum^{t}_{k=t - (t \text{ mod } q)+1} \{ f^{A}(\textbf{z}_i,k) \} & \text{if $(t \text{ mod } q)\neq0$}\\
    \end{cases} \\
    &= 
    \begin{cases}
      \alpha \cdot t & \text{if $(t \text{ mod } q)=0$}\\
      \alpha \cdot t - \alpha \cdot (t \text{ mod } q) + \sum^{t}_{k=t - (t \text{ mod } q)+1} \{ f^{A}(\textbf{z}_i,k) \} & \text{if $(t \text{ mod } q)\neq0$}\\
    \end{cases},
\end{split}
\end{equation*}
representing the number of time points with administered treatment.

We have that $\sum^{t}_{k=t - (t \text{ mod } q)+1} \{ f^{A}(\textbf{z}_i,k) \}$ is bounded between $0$ and $(t \text{ mod } q)$, while $(t \text{ mod } q)$ is bounded between $0$ and $q-1$. Therefore, for sufficiently large $t$ such that $\alpha \cdot t >> q$, we can approximate
\begin{equation*}
    \sum^{t}_{k=1} \{ \mathbb{I}(f^{A}(\textbf{z}_i,k)=1) \} \approx \alpha \cdot t.
\end{equation*}
In this case, under the additional assumption \eqref{eq:equal_sigma}, we can approximate the conditional variance of $\hat{\tau_i}$ as
\begin{equation*}
    Var(\hat{\tau_i} \mid U=u_i, \textbf{Z}=\textbf{z}_i) \approx \frac{\sigma^2(u_i)}{\alpha \cdot t} + \frac{\sigma^2(u_i)}{t-\alpha \cdot t} = \frac{\sigma^2(u_i)}{t \cdot \alpha (1-\alpha)}.
\end{equation*}

\section*{Appendix C}

Consider the basic causal model in the extreme scenario where all noise terms are equal, that is $\varepsilon_{i,k}=\varepsilon$ with $\varepsilon$ constant. We consider the intervention $\bar{A}_k=\bar{x}_k$ with $\bar{x}_k$ vector of size $k$ with elements all equal to $x$. Then, the counterfactual outcomes are constant over time 
\begin{equation*}
        Y^{\bar{a}_k=\bar{x}_k}_{i,k} \overset{\eqref{eq:carryov1}}{=} Y^{a_k=x}_{i,k} \overset{\eqref{eq:Y1}}{=} f^Y(x, U_i, \varepsilon) = Y^{a_{k_*}=x}_{i,k_*}
\end{equation*}
with $k_*$ being a generic time point.

It follows also that
\begin{equation*}
\mathbb{E}(Y^{\bar{a}_k=\bar{x}_k}_{k} \mid U=u_i) \overset{\eqref{eq:Y1}}{=} \mathbb{E}(f^Y(x, u_i, \varepsilon)) = f^Y(x, u_i, \varepsilon) = Y^{\bar{a}_k=\bar{x}_k}_{i,k}.
\end{equation*}
Therefore the individual-specific causal effect and the individual causal effect coincide in this scenario, $U\text{-}CATE_k(u_i)=ICE_{i,k}$.

Moreover, the mean of the observed outcomes for a treatment level $x$ in a fixed context $U=u_i, \textbf{Z}=\textbf{z}_i$
\begin{equation*}
\begin{split}
&\frac{1}{\sum^{t}_{k=1} \{ \mathbb{I}(f^{A}(\textbf{z}_i,k)=x) \}} \sum^{t}_{k=1} \{ \mathbb{I}(f^{A}(\textbf{z}_i,k)=x) f^Y(x, u_i, \varepsilon) \} \\
&=\frac{1}{\sum^{t}_{k=1} \{ \mathbb{I}(f^{A}(\textbf{z}_i,k)=x) \}} f^Y(x, u_i, \varepsilon) \sum^{t}_{k=1} \{ \mathbb{I}(f^{A}(\textbf{z}_i,k)=x) \} \\
&= f^Y(x, u_i, \varepsilon) = Y^{\bar{a}_k=\bar{x}_k}_{i,k}.
\end{split}  
\end{equation*}
We have, therefore, that $\hat{\tau_i}=ICE_{i,k}$.

\section*{Appendix D}

Consider the conditional expectation of the counterfactual outcome for a generic intervention $\bar{x}_k \in \{0,1\}^k$
\begin{equation*}
\begin{split}
   &\mathbb{E}(Y^{\bar{a}_k = \bar{x}_k}_{k} \mid U=u) \\
   &= \sum_{\bar{y}_{k-1}, \bar{l}_k} \mathbb{E}(Y^{\bar{a}_k = \bar{x}_k}_{k} \mid \bar{L}^{\bar{a}_{k}=\bar{x}_{k}}_{k} = \bar{l}_{k}, \bar{Y}^{\bar{a}_{k-1}=\bar{x}_{k-1}}_{k-1} = \bar{y}_{k-1}, U=u) \\
   &\mathbb{P}(\bar{L}^{\bar{a}_{k}=\bar{x}_{k}}_{k} = \bar{l}_{k}, \bar{Y}^{\bar{a}_{k-1}=\bar{x}_{k-1}}_{k-1} = \bar{y}_{k-1} \mid U=u) \\
   &= \sum_{\bar{y}_{k-1}, \bar{l}_k} \mathbb{E}(Y^{\bar{a}_k = \bar{x}_k}_{k} \mid \bar{L}^{\bar{a}_{k}=\bar{x}_{k}}_{k} = \bar{l}_{k}, \bar{Y}^{\bar{a}_{k-1}=\bar{x}_{k-1}}_{k-1} = \bar{y}_{k-1}, U=u) \\
   & \mathbb{P}(L^{\bar{a}_{k}=\bar{x}_{k}}_k=l_k \mid \bar{Y}^{\bar{a}_{k-1}=\bar{x}_{k-1}}_{k-1} = \bar{y}_{k-1}, \bar{L}^{\bar{a}_{k-1}=\bar{x}_{k-1}}_{k-1} = \bar{l}_{k-1}, U=u)\\
   &\mathbb{P}(\bar{Y}^{\bar{a}_{k-1}=\bar{x}_{k-1}}_{k-1} = \bar{y}_{k-1}, \bar{L}^{\bar{a}_{k-1}=\bar{x}_{k-1}}_{k-1} = \bar{l}_{k-1} \mid U=u) \\
   &= \sum_{\bar{y}_{k-1}, \bar{l}_k} \mathbb{E}(Y^{\bar{a}_k = \bar{x}_k}_{k} \mid \bar{L}^{\bar{a}_{k}=\bar{x}_{k}}_{k} = \bar{l}_{k}, \bar{Y}^{\bar{a}_{k-1}=\bar{x}_{k-1}}_{k-1} = \bar{y}_{k-1}, U=u) \\
   & \mathbb{P}(L^{\bar{a}_{k}=\bar{x}_{k}}_k=l_k \mid \bar{Y}^{\bar{a}_{k-1}=\bar{x}_{k-1}}_{k-1} = \bar{y}_{k-1}, \bar{L}^{\bar{a}_{k-1}=\bar{x}_{k-1}}_{k-1} = \bar{l}_{k-1}, U=u)\\
   &\mathbb{P}(Y^{\bar{a}_{k-1}=\bar{x}_{k-1}}_{k-1}=y_{k-1} \mid \bar{L}^{\bar{a}_{k-1}=\bar{x}_{k-1}}_{k-1} = \bar{l}_{k-1}, \bar{Y}^{\bar{a}_{k-2}=\bar{x}_{k-2}}_{k-2} = \bar{y}_{k-2}, U=u) \\
   &\mathbb{P}(\bar{L}^{\bar{a}_{k-1}=\bar{x}_{k-1}}_{k-1} = \bar{l}_{k-1}, \bar{Y}^{\bar{a}_{k-2}=\bar{x}_{k-2}}_{k-2} = \bar{y}_{k-2} \mid U=u)\\
   &= \sum_{\bar{y}_{k-1}, \bar{l}_k} \mathbb{E}(Y^{\bar{a}_k = \bar{x}_k}_{k} \mid \bar{L}^{\bar{a}_{k}=\bar{x}_{k}}_{k} = \bar{l}_{k}, \bar{Y}^{\bar{a}_{k-1}=\bar{x}_{k-1}}_{k-1} = \bar{y}_{k-1}, U=u) \\
   & \mathbb{P}(L^{\bar{a}_{k}=\bar{x}_{k}}_k=l_k \mid \bar{Y}^{\bar{a}_{k-1}=\bar{x}_{k-1}}_{k-1} = \bar{y}_{k-1}, \bar{L}^{\bar{a}_{k-1}=\bar{x}_{k-1}}_{k-1} = \bar{l}_{k-1}, U=u)\\
   &\mathbb{P}(Y^{\bar{a}_{k-1}=\bar{x}_{k-1}}_{k-1}=y_{k-1} \mid \bar{L}^{\bar{a}_{k-1}=\bar{x}_{k-1}}_{k-1} = \bar{l}_{k-1}, \bar{Y}^{\bar{a}_{k-2}=\bar{x}_{k-2}}_{k-2} = \bar{y}_{k-2}, U=u) \\
   &\mathbb{P}(L^{\bar{a}_{k-1}=\bar{x}_{k-1}}_{k-1} = l_{k-1} \mid \bar{Y}^{\bar{a}_{k-2}=\bar{x}_{k-2}}_{k-2} = \bar{y}_{k-2}, \bar{L}^{\bar{a}_{k-2}=\bar{x}_{k-2}}_{k-2} = \bar{l}_{k-2}, U=u) \\
   &\mathbb{P}(\bar{Y}^{\bar{a}_{k-2}=\bar{x}_{k-2}}_{k-2} = \bar{y}_{k-2}, \bar{L}^{\bar{a}_{k-2}=\bar{x}_{k-2}}_{k-2} = \bar{l}_{k-2} \mid U=u).
    \end{split}
\end{equation*}

Continuing iteratively, we find that 
\begin{equation*}
    \begin{split}
    &\mathbb{E}(Y^{\bar{a}_k = \bar{x}_k}_{k} \mid U=u) = \sum_{\bar{y}_{k-1}, \bar{l}_k} \mathbb{E}(Y^{\bar{a}_k = \bar{x}_k}_{k} \mid \bar{L}^{\bar{a}_{k}=\bar{x}_{k}}_{k} = \bar{l}_{k}, \bar{Y}^{\bar{a}_{k-1}=\bar{x}_{k-1}}_{k-1} = \bar{y}_{k-1}, U=u) \\
   & \mathbb{P}(L^{\bar{a}_{k}=\bar{x}_{k}}_k=l_k \mid \bar{Y}^{\bar{a}_{k-1}=\bar{x}_{k-1}}_{k-1} = \bar{y}_{k-1}, \bar{L}^{\bar{a}_{k-1}=\bar{x}_{k-1}}_{k-1} = \bar{l}_{k-1}, U=u)\\
   &\prod^{k-1}_{m=1} \{ \mathbb{P}(Y^{\bar{a}_{m}=\bar{x}_{m}}_{m}=y_{m} \mid \bar{L}^{\bar{a}_{m}=\bar{x}_{m}}_{m} = \bar{l}_{m}, \bar{Y}^{\bar{a}_{m-1}=\bar{x}_{m-1}}_{m-1} = \bar{y}_{m-1}, U=u) \\
   &\mathbb{P}(L^{\bar{a}_{m}=\bar{x}_{m}}_{m} = l_{m} \mid \bar{Y}^{\bar{a}_{m-1}=\bar{x}_{m-1}}_{m-1} = \bar{y}_{m-1}, \bar{L}^{\bar{a}_{m-1}=\bar{x}_{m-1}}_{m-1} = \bar{l}_{m-1}, U=u) \}.
    \end{split}
\end{equation*}

The formula can be further simplified for the specific causal model we consider in Section \ref{sec:relaxed_scenario}. From the DAG in Figure \ref{fig:dgm3} it follows that $L^{\bar{a}_k}_k$ is independent of any other variable that is non-descendent of $L^{\bar{a}_k}_k$ when conditioning on $Y^{\bar{a}_{k-1}}_{k-1}, L^{\bar{a}_{k-1}}_{k-1}$ and $U$. Similarly, it follows from the DAG also that $Y^{\bar{a}_k}_k$ is independent from any non-descendent variable once conditioned on $L^{\bar{a}_k}_{k}, Y^{\bar{a}_{k-1}}_{k-1}, L^{\bar{a}_{k-1}}_{k-1}$ and $U$. Therefore we have
\begin{equation*}
    \begin{split}
    &\mathbb{E}(Y^{\bar{a}_k = \bar{x}_k}_{k} \mid U=u) =\\
    &\sum_{\bar{y}_{k}, \bar{l}_k} y_k \, \mathbb{P}(Y^{\bar{a}_k = \bar{x}_k}_{k}=y_k \mid L^{\bar{a}_{k}=\bar{x}_{k}}_{k} = l_{k}, Y^{\bar{a}_{k-1}=\bar{x}_{k-1}}_{k-1} = y_{k-1}, L^{\bar{a}_{k-1}=\bar{x}_{k-1}}_{k-1} = l_{k-1}, U=u) \\
   & \mathbb{P}(L^{\bar{a}_{k}=\bar{x}_{k}}_k=l_k \mid Y^{\bar{a}_{k-1}=\bar{x}_{k-1}}_{k-1} = y_{k-1}, L^{\bar{a}_{k-1}=\bar{x}_{k-1}}_{k-1} = l_{k-1}, U=u)\\
   &\prod^{k-1}_{m=1} \{ \mathbb{P}(Y^{\bar{a}_m = \bar{x}_m}_{m}=y_m \mid L^{\bar{a}_{m}=\bar{x}_{m}}_{m} = l_{m}, Y^{\bar{a}_{m-1}=\bar{x}_{m-1}}_{m-1} = y_{m-1}, L^{\bar{a}_{m-1}=\bar{x}_{m-1}}_{m-1} = l_{m-1}, U=u) \\
   & \mathbb{P}(L^{\bar{a}_{m}=\bar{x}_{m}}_m=l_m \mid Y^{\bar{a}_{m-1}=\bar{x}_{m-1}}_{m-1} = y_{m-1}, L^{\bar{a}_{m-1}=\bar{x}_{m-1}}_{m-1} = l_{m-1}, U=u) \}\\
   &=\sum_{\bar{y}_{k}, \bar{l}_k} y_k \, \prod^{k}_{m=1} \{ \mathbb{P}(Y^{\bar{a}_m = \bar{x}_m}_{m}=y_m \mid L^{\bar{a}_{m}=\bar{x}_{m}}_{m} = l_{m}, Y^{\bar{a}_{m-1}=\bar{x}_{m-1}}_{m-1} = y_{m-1}, L^{\bar{a}_{m-1}=\bar{x}_{m-1}}_{m-1} = l_{m-1}, U=u) \\   & \mathbb{P}(L^{\bar{a}_{m}=\bar{x}_{m}}_m=l_m \mid Y^{\bar{a}_{m-1}=\bar{x}_{m-1}}_{m-1} = y_{m-1}, L^{\bar{a}_{m-1}=\bar{x}_{m-1}}_{m-1} = l_{m-1}, U=u) \}.
    \end{split}
\end{equation*}

We define
\begin{equation*}
\begin{split}
&k^{\textbf{z}}_{x} \in \{k_* \in \{1,...,t\} \mid f^{A}(\textbf{z},k_*)=x\}\\
&k^{\textbf{z}}_{x_1x_2} \in \{k_* \in \{1,...,t\} \mid \mathbb{I}(f^{A}(\textbf{z},k_*)=x_2)\mathbb{I}(f^{A}(\textbf{z},k_*-1)=x_1)=1\}
\end{split}
\end{equation*}
so that, for example, for a specific treatment schedule $\textbf{z}$, $k^{\textbf{z}}_{00}$ represents a time where control was administered twice in a row, and $k^{\textbf{z}}_1$ represents a time where treatment was administered.

 We can define
\begin{equation*}
g_{k^{\textbf{z}}_{x}}^L(l \mid y', l', u) =   \mathbb{P}(L_{k^{\textbf{z}}_{x}}=l \mid A_{k^{\textbf{z}}_{x}}=x, Y_{k^{\textbf{z}}_{x}-1}=y', L_{k^{\textbf{z}}_{x}-1}=l', U=u, \textbf{Z}=\textbf{z})
\end{equation*}
and

\begin{equation*}
\begin{split}
&g_{k^{\textbf{z}}_{x_1x_2}}^Y(y \mid l, y', l', u) = \\
&\mathbb{P}(Y_{k^{\textbf{z}}_{x_1x_2}}= y \mid L_{k^{\textbf{z}}_{x_1x_2}}=l, A_{k^{\textbf{z}}_{x_1x_2}}=x_2, Y_{k^{\textbf{z}}_{x_1x_2}-1}=y', L_{k^{\textbf{z}}_{x_1x_2}-1}=l', A_{k^{\textbf{z}}_{x_1x_2}-1}=x_1, U=u, \textbf{Z}=\textbf{z})
\end{split}
\end{equation*}
When $k^{\textbf{z}}_{x_1x_2}\geq2$ we have
\begin{equation*}
\begin{split}
&g_{k^{\textbf{z}}_{x_1x_2}}^Y(y \mid l, y',  l',  u) \overset{\eqref{eq:consistency2}}{=} \\
&\mathbb{P}(Y^{\bar{a}_{k^{\textbf{z}}_{x_1x_2}} = \bar{v}}_{k^{\textbf{z}}_{x_1x_2}}= y \mid L^{\bar{a}_{k^{\textbf{z}}_{x_1x_2}} = \bar{v}}_{k^{\textbf{z}}_{x_1x_2}}=l, A_{k^{\textbf{z}}_{x_1x_2}}=x_2, Y^{\bar{a}_{k^{\textbf{z}}_{x_1x_2}} = \bar{v}}_{k^{\textbf{z}}_{x_1x_2}-1}=y',  L^{\bar{a}_{k^{\textbf{z}}_{x_1x_2}} = \bar{v}}_{k^{\textbf{z}}_{x_1x_2}-1}=l',  A_{k^{\textbf{z}}_{x_1x_2}-1}=x_1, U=u, \textbf{Z}=\textbf{z})
\end{split}
\end{equation*}
with $\bar{v}=(f^{A}(\textbf{z},1),...,f^{A}(\textbf{z},k^{\textbf{z}}_{x_1x_2}))$ representing the natural treatment sequence until time $k^{\textbf{z}}_{x_1x_2}$ when $\textbf{Z}=\textbf{z}$. Furthermore, since $Y^{\bar{a}_{k^{\textbf{z}}_{x_1x_2}}}_{k^{\textbf{z}}_{x_1x_2}} \independent \bar{A}_{k^{\textbf{z}}_{x_1x_2}}, \textbf{Z}$ conditional on the other variables, we have 
\begin{equation*}
g_{k^{\textbf{z}}_{x_1x_2}}^Y(y \mid l, y',  l',  u)=
\mathbb{P}(Y^{\bar{a}_{k^{\textbf{z}}_{x_1x_2}} = \bar{v}}_{k^{\textbf{z}}_{x_1x_2}}= y \mid L^{\bar{a}_{k^{\textbf{z}}_{x_1x_2}} = \bar{v}}_{k^{\textbf{z}}_{x_1x_2}}=l,  Y^{\bar{a}_{k^{\textbf{z}}_{x_1x_2}} = \bar{v}}_{k^{\textbf{z}}_{x_1x_2}-1}=y',  L^{\bar{a}_{k^{\textbf{z}}_{x_1x_2}} = \bar{v}}_{k^{\textbf{z}}_{x_1x_2}-1}=l',  U=u).
\end{equation*}
Since $k^{\textbf{z}}_{x_1x_2}$ is defined as the time point in which treatments $x_1,x_2$ have been administered consecutively under $\textbf{z}$, $\bar{v}$ can be rewritten as $\bar{v}=( f^{A}(\textbf{z},1),...,f^{A}(\textbf{z},k^{\textbf{z}}_{x_1x_2}-2),x_1,x_2)$.
Therefore, due to the assumption of stationarity \eqref{eq:stationarity2}, we have 
\begin{equation*}
\begin{split}
&g_{k^{\textbf{z}}_{x_1x_2}}^Y(y \mid l, y',  l',  u) \overset{\eqref{eq:stationarity2}}{=}\\
&\mathbb{P}(Y^{\bar{a}_{k_*}=(\bar{v}', x_1,x_2)}_{k_*}=y \mid L^{\bar{a}_{k_*}=(\bar{v}',x_1,x_2)}_{k_*}=l, Y^{\bar{a}_{k_*-1}=(\bar{v}',x_1)}_{k_*-1}=y', L^{\bar{a}_{k_*-1}=(\bar{v}',x_1)}_{k_*-1}=l', U=u)
\end{split}
\end{equation*}
with $k_* \in \{1,...,t\}$ being a generic time point and $\bar{v}' \in \{0,1\}^{k_*-2}$.
This means that, conditional on $U=u, \textbf{Z}=\textbf{z}$, all time points $k^{\textbf{z}}_{x_1x_2}$ for which treatment $x_2$ was administered after treatment $x_1$ have the same conditional distribution of the outcome given its direct causes, which corresponds to the conditional distribution of the counterfactual outcome under any intervention in which $x_1,x_2$ are assigned. Since the stationarity condition also applies to the first time point, an analogous argument holds when $k^{\textbf{z}}_{x_1x_2}=1$, interpreting the history preceding the first time point as the fixed initial conditions. Following a similar reasoning, due to consistency \eqref{eq:consistency2} and stationarity \eqref{eq:stationarityL}, 
\begin{equation*}
g_{k^{\textbf{z}}_{x}}^L(l \mid y', l', u)=
\mathbb{P}(L^{\bar{a}_{k_*}=(\bar{v}', x)}_{k_*}=l \mid Y^{\bar{a}_{k_*-1}=\bar{v}'}_{k_*-1}=y', L^{\bar{a}_{k_*-1}=\bar{v}'}_{k_*-1}=l', U=u).
\end{equation*}
Condition \eqref{eq:Z2} guarantees that, when $t>q$, for every treatment schedule $\textbf{Z}$ there exists at least one time point in which the treatment sequence $x_1,x_2$ is observed for any $x_1,x_2 \in \{0,1\}$. Condition \eqref{eq:positivity} instead guarantees that, for every $U=u$ existing in the population, the conditional probabilities are well defined.

We can therefore rewrite the conditional expected counterfactual outcome as 
\begin{equation*}
\begin{split}
&\mathbb{E}(Y^{\bar{a}_k = \bar{x}_k}_{k} \mid U=u) =\\
&\sum_{\bar{y}_{k}, \bar{l}_k} y_k \,
\prod^{k}_{m=1} g_{k^{\textbf{z}}_{x_{m-1}x_{m}}}^Y(y_{m} \mid l_{m}, y_{m-1}, l_{m-1},  u) \, g_{k^{\textbf{z}}_{x_m}}^L(l_m \mid y_{m-1}, l_{m-1}, u) \overset{\eqref{eq:theta}}{=}\theta_k(\bar{x}_k,u,\textbf{z}).
\end{split}
\end{equation*}
This result is a special case of the g-formula reported in Robins et al.\ 1999. Specifically, Robins and colleagues discuss identification of subject-specific effects leveraging stationarity in the “Conditional Methods” section of their article. In their work, Robins and colleagues do not invoke stationarity of the counterfactual distributions (our Assumptions \eqref{eq:stationarityL} and \eqref{eq:stationarity2}), but a weaker version of stationarity involving only observed conditional distributions. Indeed, when there is a positive probability of receiving treatment at any time point, stationarity of the counterfactuals is not needed for identification, and a weaker stationarity assumption of the observed conditional distributions is sufficient for the estimation.

More formally, assume that the following positivity condition holds
\begin{equation*}
    \mathbb{P}(\bar{Y}=\bar{y}, \bar{L}=\bar{l}, \bar{A}=\bar{x}_t \mid U=u) > 0
\end{equation*}
for every $\bar{y} \in \mathcal{Y}^t$, every $\bar{l} \in \mathcal{L}^t$, every $\bar{x}_t \in \{0,1\}^t$, and every $u \in \mathcal{U}$.

We have already seen that
\begin{equation*}
    \begin{split}
    &\mathbb{E}(Y^{\bar{a}_k = \bar{x}_k}_{k} \mid U=u) =\\
    &\sum_{\bar{y}_{k}, \bar{l}_k} y_k \, \prod^{k}_{m=1} \{ \mathbb{P}(Y^{\bar{a}_m = \bar{x}_m}_{m}=y_m \mid L^{\bar{a}_{m}=\bar{x}_{m}}_{m} = l_{m}, Y^{\bar{a}_{m-1}=\bar{x}_{m-1}}_{m-1} = y_{m-1}, L^{\bar{a}_{m-1}=\bar{x}_{m-1}}_{m-1} = l_{m-1}, U=u) \\
   & \mathbb{P}(L^{\bar{a}_{m}=\bar{x}_{m}}_m=l_m \mid Y^{\bar{a}_{m-1}=\bar{x}_{m-1}}_{m-1} = y_{m-1}, L^{\bar{a}_{m-1}=\bar{x}_{m-1}}_{m-1} = l_{m-1}, U=u) \}.
    \end{split}
\end{equation*}
Since from the DAG in Figure \ref{fig:dgm3} we have $Y^{\bar{a}_k}_k \independent \bar{A}_k | L^{\bar{a}_k}_k, Y^{\bar{a}_{k-1}}_{k-1}, L^{\bar{a}_{k-1}}_{k-1},U$ and $L^{\bar{a}_k}_k \independent \bar{A}_k | Y^{\bar{a}_{k-1}}_{k-1}, L^{\bar{a}_{k-1}}_{k-1},U$, we can rewrite 
\begin{equation*}
    \begin{split}
    &\mathbb{E}(Y^{\bar{a}_k = \bar{x}_k}_{k} \mid U=u) =\\
    &\sum_{\bar{y}_{k}, \bar{l}_k} y_k \, \prod^{k}_{m=1} \{ \mathbb{P}(Y^{\bar{a}_m = \bar{x}_m}_{m}=y_m \mid \bar{A}_m=\bar{x}_m, L^{\bar{a}_{m}=\bar{x}_{m}}_{m} = l_{m}, Y^{\bar{a}_{m-1}=\bar{x}_{m-1}}_{m-1} = y_{m-1}, L^{\bar{a}_{m-1}=\bar{x}_{m-1}}_{m-1} = l_{m-1}, U=u) \\
   & \mathbb{P}(L^{\bar{a}_{m}=\bar{x}_{m}}_m=l_m \mid \bar{A}_m=\bar{x}_m, Y^{\bar{a}_{m-1}=\bar{x}_{m-1}}_{m-1} = y_{m-1}, L^{\bar{a}_{m-1}=\bar{x}_{m-1}}_{m-1} = l_{m-1}, U=u) \}.
    \end{split}
\end{equation*}
By consistency \eqref{eq:consistency2}, we can further rewrite 
\begin{equation*}
    \begin{split}
    &\mathbb{E}(Y^{\bar{a}_k = \bar{x}_k}_{k} \mid U=u) =\\
    &\sum_{\bar{y}_{k}, \bar{l}_k} y_k \, \prod^{k}_{m=1} \{ \mathbb{P}(Y_{m}=y_m \mid \bar{A}_m=\bar{x}_m, L_{m} = l_{m}, Y_{m-1} = y_{m-1}, L_{m-1} = l_{m-1}, U=u) \\
   & \mathbb{P}(L_m=l_m \mid \bar{A}_m=\bar{x}_m, Y_{m-1} = y_{m-1}, L_{m-1} = l_{m-1}, U=u) \}\\
   &=\sum_{\bar{y}_{k}, \bar{l}_k} y_k \, \prod^{k}_{m=1} \{ \mathbb{P}(Y_{m}=y_m \mid A_m=x_m, A_{m-1}=x_{m-1}, L_{m} = l_{m}, Y_{m-1} = y_{m-1}, L_{m-1} = l_{m-1}, U=u) \\
   & \mathbb{P}(L_m=l_m \mid A_m=x_m, Y_{m-1} = y_{m-1}, L_{m-1} = l_{m-1}, U=u) \},
    \end{split}
\end{equation*}
where the last step comes from the fact that conditioning on all direct causes of a variable, the variable is independent of all other non-descendant variables. This completes the identification argument, without invoking stationarity of the counterfactuals. As argued in Robins et al.\ 1999, if the conditional distributions of $Y_k$ and $L_k$ given their direct causes do not change over time, it is possible to estimate $\mathbb{E}(Y^{\bar{a}_k = \bar{x}_k}_{k} \mid U=u)$ using data from a single unit $i$. That is, if there is stationarity of the observed conditional distributions, it is then possible to estimate the $U\text{-}CATE_k(u_i)$.

\section*{Appendix E}

We consider a causal model compatible with the DAG in Figure \ref{fig:dgm1} under a FFRCISTG interpretation, with treatment assignment according to Equation \eqref{eq:A1}. Under this data generating mechanism, there is no carryover effect of the treatment (condition \eqref{eq:carryov1} holds).

Further, suppose that individuals are randomized only to treatment schedules

\begin{equation*} \mathcal{Z} \subseteq \bigg\{ \textbf{z}=(z_1,...,z_q) \bigg\rvert \sum^{t}_{k=1} \{ f^{A}(\textbf{z},k)\} = t \cdot \alpha^* \bigg\}
\end{equation*}
with $0<\alpha^*<1$, and such that
$$\mathbb{P}(f^{A}(\textbf{Z},k)=1)=\alpha^*$$ for every $k \in \{1,...,t\}$. For example, this treatment assignment mechanism is compatible with a 28-day N-of-1 trial with daily measurements, where the participant is randomized with equal probability to a treatment schedule in $\mathcal{Z} = \{(\bar{0}_7 ,\bar{1}_7),(\bar{1}_7 ,\bar{0}_7)\}$. We will not require any additional assumption.

We will use that
\begin{equation*}
    \begin{split}
        &\frac{1}{\sum^{t}_{k=1} \{ \mathbb{I}(A_{i,k}=x) \}} \sum^{t}_{k=1} \{ \mathbb{I}(A_{i,k}=x) Y_{i,k} \} \\
        &\overset{\eqref{eq:A1}}{=} \frac{1}{\sum^{t}_{k=1} \{ \mathbb{I}(f^{A}(\textbf{Z}_i,k)=x) \}} \sum^{t}_{k=1} \{ \mathbb{I}(f^{A}(\textbf{Z}_i,k)=x) Y_{i,k} \}\\
        &\overset{\eqref{eq:consistency} \eqref{eq:carryov1}}{=} \frac{1}{\sum^{t}_{k=1} \{ \mathbb{I}(f^{A}(\textbf{Z}_i,k)=x) \}} \sum^{t}_{k=1} \{ \mathbb{I}(f^{A}(\textbf{Z}_i,k)=x) Y^{\bar{a}_k=\bar{x}_k}_{i,k} \}.
    \end{split}
\end{equation*}

Consider the following expectation for $x=1$ conditioned on $U=u_i, \bar{\varepsilon}=\bar{\varepsilon}_i$,
\begin{equation*}
\begin{split}
    &\mathbb{E} \bigg(  \frac{1}{\sum^{t}_{k=1} \{ \mathbb{I}(f^{A}(\textbf{Z},k)=1) \}} \sum^{t}_{k=1} \{ \mathbb{I}(f^{A}(\textbf{Z},k)=1) Y^{\bar{a}_k=\bar{1}_k}_k \} \bigg| U=u_i, \bar{\varepsilon}=\bar{\varepsilon}_i \bigg) \\ 
    &=\mathbb{E} \bigg(  \frac{1}{t \cdot \alpha^*} \sum^{t}_{k=1} \{ \mathbb{I}(f^{A}(\textbf{Z},k)=1) Y^{\bar{a}_k=\bar{1}_k}_k \} \bigg| U=u_i, \bar{\varepsilon}=\bar{\varepsilon}_i \bigg) \\ 
    &=\frac{1}{t \cdot \alpha^*} \sum^{t}_{k=1} \{ \mathbb{E} (\mathbb{I}(f^{A}(\textbf{Z},k)=1) Y^{\bar{a}_k=\bar{1}_k}_k \mid U=u_i, \bar{\varepsilon}=\bar{\varepsilon}_i ) \}  \\     &\overset{(Y^{\bar{a}_k}_k \independent \textbf{Z} \mid U, \bar{\varepsilon})}{=} \frac{1}{t \cdot \alpha^*} \sum^{t}_{k=1} \{ \mathbb{E} (\mathbb{I}(f^{A}(\textbf{Z},k)=1) \mid U=u_i, \bar{\varepsilon}=\bar{\varepsilon}_i ) \cdot \mathbb{E} (Y^{\bar{a}_k=\bar{1}_k}_k \mid U=u_i, \bar{\varepsilon}=\bar{\varepsilon}_i ) \}\\
    &\overset{(\textbf{Z}\independent U, \bar{\varepsilon})}{=} \frac{1}{t \cdot \alpha^*} \sum^{t}_{k=1} \{ \mathbb{E} (\mathbb{I}(f^{A}(\textbf{Z},k)=1)) \cdot \mathbb{E} (Y^{\bar{a}_k=\bar{1}_k}_k \mid U=u_i, \bar{\varepsilon}=\bar{\varepsilon}_i ) \}  \\
    &= \frac{1}{t \cdot \alpha^*} \sum^{t}_{k=1} \{ \mathbb{P} (f^{A}(\textbf{Z},k)=1) \cdot \mathbb{E} (Y^{\bar{a}_k=\bar{1}_k}_k \mid U=u_i, \bar{\varepsilon}=\bar{\varepsilon}_i ) \}  \\
    &= \frac{\alpha^*}{t \cdot \alpha^*} \sum^{t}_{k=1} \{ Y^{\bar{a}_k=\bar{1}_k}_{i,k}  \}  = \frac{1}{t} \sum^{t}_{k=1} \{ Y^{\bar{a}_k=\bar{1}_k}_{i,k} \}.
\end{split}
\end{equation*}

Analogously, we can show that 
\begin{equation*}
\begin{split}
    &\mathbb{E} \bigg(  \frac{1}{\sum^{t}_{k=1} \{ \mathbb{I}(f^{A}(\textbf{Z},k)=0) \}} \sum^{t}_{k=1} \{ \mathbb{I}(f^{A}(\textbf{Z},k)=0) Y^{\bar{a}_k=\bar{0}_k}_k \} \bigg| U=u_i, \bar{\varepsilon}=\bar{\varepsilon}_i \bigg) \\ 
    &= \frac{1}{t} \sum^{t}_{k=1} \{ Y^{\bar{a}_k=\bar{0}_k}_{i,k} \}.
\end{split}
\end{equation*}
It follows that 
\begin{equation*}
    \mathbb{E} (\hat{\tau_i} \mid U=u_i, \bar{\varepsilon}=\bar{\varepsilon}_i) = \frac{1}{t} \sum^{t}_{k=1} \{ Y^{\bar{a}_k=\bar{1}_k}_{i,k} - Y^{\bar{a}_k=\bar{0}_k}_{i,k}\}\overset{\eqref{eq:ice}}{=} \frac{1}{t} \sum^{t}_{k=1} ICE_{i,k}.  
\end{equation*}

\section*{Appendix F}

Consider the average causal effect at time $k$, we have

\begin{equation*}
\begin{split}
    ACE_k &= \mathbb{E}(Y^{\bar{a}_k = \bar{1}_k}_{k}) - \mathbb{E}(Y^{\bar{a}_k = \bar{0}_k}_{k} ) \overset{\eqref{eq:carryov1}}{=} \mathbb{E} ( Y^{a_k = 1}_{k} - Y^{a_k = 0}_{k} )\\
    & = \mathbb{E} \bigg( \mathbb{E} ( Y^{a_k = 1}_{k} - Y^{a_k = 0}_{k} \mid U) \bigg) 
    \overset{\eqref{eq:stationarity1_expectation}}{=} \mathbb{E} \bigg( \mathbb{E} ( Y^{a_{k_*} = 1}_{k_*} - Y^{a_{k_*} = 0}_{k_*} \mid U) \bigg)\\
    &= \mathbb{E} ( Y^{a_{k_*} = 1}_{k_*} - Y^{a_{k_*} = 0}_{k_*}) \overset{\eqref{eq:carryov1}}{=} ACE_{k_*}
\end{split}
\end{equation*}

with $k_*$ being a generic time point. Therefore, the $ACE_k$ is constant over time.

\subsection*{Unbiasedness}

Consider the marginal expected value of $\frac{1}{n} \sum^n_{i=1} \hat{\tau_i}$
\begin{equation*}
    \begin{split}
        \mathbb{E} \bigg( \frac{1}{n} \sum^n_{i=1} \hat{\tau_i} \bigg) &=  \mathbb{E} \bigg( \mathbb{E} \bigg( \frac{1}{n} \sum^n_{i=1} \hat{\tau_i} \bigg| U, \textbf{Z} \bigg) \bigg) = \frac{1}{n} \sum^n_{i=1} \{ \mathbb{E} ( \mathbb{E} ( \hat{\tau_i} \mid U, \textbf{Z} ) ) \} \\
        &\overset{(B)}{=} \frac{1}{n} \sum^n_{i=1} \{ \mathbb{E} ( \tau(U,\textbf{Z})  ) \} \overset{(A)}{=} \frac{1}{n} \sum^n_{i=1} \{ \mathbb{E} ( U\text{-}CATE_k (U) ) \}  \\
        &\overset{\eqref{eq:ucate1}}{=} \frac{1}{n} \sum^n_{i=1} \{\mathbb{E}( \mathbb{E}(Y^{\bar{a}_k = \bar{1}_k}_{k} \mid U) - \mathbb{E}(Y^{\bar{a}_k = \bar{0}_k}_{k} \mid U) ) \} \\
        &= \frac{1}{n} \sum^n_{i=1} \{ \mathbb{E}(Y^{\bar{a}_k = \bar{1}_k}_{k}) - \mathbb{E}(Y^{\bar{a}_k = \bar{0}_k}_{k} ) \} \overset{\eqref{eq:ACE}}{=} \frac{1}{n} \sum^n_{i=1} \{ACE_k\} = ACE_k
    \end{split}
\end{equation*}

\subsection*{Consistency}

Because the individuals are
i.i.d. draws from the population, the estimators 
$\hat{\tau_i}$ are i.i.d. random variables. Moreover,
from the unbiasedness result above, we have $\mathbb{E}(\hat{\tau_i})=ACE_k$.
Therefore, as $n$ goes to infinity,
\begin{equation*}
    \frac{1}{n}\sum_{i=1}^n \hat{\tau}_i
    \xrightarrow{P}
    \mathbb{E}(\hat{\tau}_i)
    =
    ACE_k
\end{equation*}
by the weak law of large numbers.

Therefore the mean of $\hat{\tau_i}$ is a consistent estimator for the $ACE_k$ for $n$ that goes to infinity.

\section*{Appendix G}

Consider

\begin{equation*}
\frac{1}{n} \sum^{n}_{i=1}( \hat{\theta}^{1}_{i,k}-\hat{\theta}^{0}_{i,k} ) =  \sum^{ }_{\textbf{z} \in \mathcal{Z}, u \in \mathcal{U}} \bigg\{ \frac{1}{n} \sum^{n}_{i=1} (\hat{\theta}^{1}_{i,k}-\hat{\theta}^{0}_{i,k}) \mathbb{I}(U_i=u, \textbf{Z}_i=\textbf{z}) \bigg\}
\end{equation*}

We defined $\hat{\theta}^{1}_{i,k}-\hat{\theta}^{0}_{i,k}$ such that, for each fixed $k$, it converges in probability towards $U\text{-}CATE_k(u_i)$,  conditional on $U=u_i$ and $\textbf{Z}=\textbf{z}_i$, when $t$ goes to infinity. Therefore, for fixed $n$ and fixed $k$, as $t$ goes to infinity, we have
\begin{equation*}
\begin{split}
\sum^{ }_{\textbf{z} \in \mathcal{Z}, u \in \mathcal{U}} \bigg\{ \frac{1}{n} \sum^{n}_{i=1} (\hat{\theta}^{1}_{i,k}-\hat{\theta}^{0}_{i,k}) \mathbb{I}(U_i=u, \textbf{Z}_i=\textbf{z}) \bigg\}  \xrightarrow{P} \\
  \sum^{ }_{\textbf{z} \in \mathcal{Z}, u \in \mathcal{U}} \bigg\{ \frac{1}{n} \sum^{n}_{i=1} U\text{-}CATE_k(u) \mathbb{I}(U_i=u, \textbf{Z}_i=\textbf{z}) \bigg\}.    
\end{split}
\end{equation*}

After taking the limit as $t$ goes to infinity for fixed $n$, let $n$ go to infinity. Because the individuals are i.i.d., the weak law of large numbers gives
\begin{equation*}
\begin{split}
  &\sum^{ }_{\textbf{z} \in \mathcal{Z}, u \in \mathcal{U}} \bigg\{ \frac{1}{n} \sum^{n}_{i=1} U\text{-}CATE_k(u) \mathbb{I}(U_i=u, \textbf{Z}_i=\textbf{z}) \bigg\} 
    \xrightarrow{P} \\
   &\sum^{ }_{\textbf{z} \in \mathcal{Z}, u \in \mathcal{U}} U\text{-}CATE_k(u) \mathbb{P}(U=u, \textbf{Z}=\textbf{z})\\
   &=\sum^{ }_{\textbf{z} \in \mathcal{Z}, u \in \mathcal{U}} \mathbb{E} ( Y^{\bar{a}_{k} = \bar{1}_k}_{k} - Y^{\bar{a}_{k} = \bar{0}_k}_{k} \mid U=u) \mathbb{P}(U=u, \textbf{Z}=\textbf{z})\\
   &=\mathbb{E}(Y^{\bar{a}_k = \bar{1}_k}_{k} - Y^{\bar{a}_k = \bar{0}_k}_{k} ) =ACE_k.    
\end{split}
\end{equation*}

\section*{Appendix H}

In this Appendix, we will consider adjustment for an unobserved time-trend-inducing variable $\bar L$. Throughout, we will assume that were $\bar L$ observed, the stationarity and positivity conditions to identify the U-CATE via the g-formula \eqref{eq:theta} would be satisfied conditional on relevant $L$ history and possibly other observed covariates. 

We consider the setting where past treatment is not an ancestor of any non-treatment input to the stationary outcome law. Specifically, we focus on the setting where the only non-treatment input in the outcome law whose distribution changes over time is the unobserved process $L$. The data generating process considered in Section \ref{sec:time_trend} falls under this category. Yet, our reasoning can be easily adapted to a setting where the stationary outcome law additionally requires an observed
treatment-unaffected input.

We will describe two approaches to adjustment. The first approach uses time as a proxy in regression models for laws that marginalize over the latent $L$. The second approach fits a latent variable model that includes a model specification for the latent $L$. 

\subsection*{Time as a proxy}\label{sec:case1atimeasproxy}
Let
\[
S_k=S(k,\bz)
\]
be a declared deterministic summary of time (e.g. spline basis, Fourier transform, day of week indicator, etc). Let
\[
\Lambda_k=(L_k,L_{k-1},\ldots,L_{k-p})
\]
denote exactly the current and lagged values of $L$ needed for the
stationary outcome law at time $k$. Let
\[
V_k=V(\bar A_k,k,\bZ),
\qquad
v_k(\bar x)=V(\bar x_k,k,\bZ),
\]
be the summary of treatment history affecting $Y_k$. Thus $V_k$ may contain
current treatment, lagged treatments, cumulative treatment, time since a
treatment change, or any other deterministic function of the assigned
treatment sequence.

\begin{assumption}[Exogeneity: No arrows from treatment into $L$]
\label{ass:L-invariance}
For every intervention $\bar x$,
\begin{equation}
\Lambda_k^{\bar x}=\Lambda_k .
\label{eq:L-invariance}
\end{equation}
\end{assumption}

\begin{assumption}[Full-data stationarity]
\label{ass:full-1a}
For every time,
intervention, and $\lambda$,
\begin{equation}
\mathbb{P}\!\left(Y_k^{\bar x}=y\mid
\Lambda_k^{\bar x}=\lambda,U=u,\bZ=\bz\right)
=q^{\mathrm{full}}\!\left(
y\mid v_k(\bar x),\lambda,u,\bz\right).
\label{eq:full-1a}
\end{equation}
The same kernel governs the corresponding conditional law under the
observed treatment schedule.
\end{assumption}

Equation \eqref{eq:full-1a} gives the full data outcome
law and states the stationarity condition of the main text. Once the
treatment-history summary and the relevant values of $L$ are fixed, the
right-hand side has no additional dependence on the time index $k$.

Because $\Lambda_k$ is unobserved, define the marginal outcome law by
\begin{equation}
q_k^{\mathrm{red}}(y\mid v,u,\bz)
=\int q^{\mathrm{full}}(y\mid v,\lambda,u,\bz)\,
dF_{\Lambda_k\mid U,\bZ}(\lambda\mid u,\bz).
\label{eq:reduced-1a}
\end{equation}
Unlike the full-data kernel, this marginal or reduced kernel may vary with $k$.
Even if \eqref{eq:full-1a} is stationary, the distribution being
integrated in \eqref{eq:reduced-1a} may differ across times.

\begin{assumption}[Time-summary model for the marginal/reduced distribution]
\label{ass:reduced-1a}
For a declared model class $\mathcal Q^S$, there is a kernel
$q^{\mathrm{red},S}\in\mathcal Q^S$ such that
\begin{equation}
q_k^{\mathrm{red}}(y\mid v,u,\bz)
=q^{\mathrm{red},S}(y\mid v,S_k,u,\bz)
\label{eq:summary-1a}
\end{equation}
at every observational or intervention-relevant value of $(v,S_k)$.
\end{assumption}

Assumption~\ref{ass:reduced-1a} states that $S_k$ indexes the changes across time in the law
that remains after $\Lambda_k$ has been integrated out in such a way that a model from $\mathcal Q^S$ is well specified. For example, $S_k$ might be a day of week indicator, in which case $Q^S$ need not necessarily impose parametric assumptions if there is sufficient revisitation to the values $S$ takes. Alternatively, perhaps $S_k$ takes a different value for each $k$ (e.g. a spline basis, or $k$ itself), in which case $\mathcal Q^S$ would need to be a restrictive parametric (e.g. linear) model.

\begin{assumption}[Consistency and support]
\label{ass:common}
The observed variables equal their counterfactual values under the
observed treatment schedule. Every value of $(v,S_k,u,\bz)$ is supported by the observed schedule, or the
required extrapolation within the declared model class is stated as an
additional assumption.
\end{assumption}

Define
\begin{equation}
m^{\mathrm{red},S}(v,s,u,\bz)
=\int y\,q^{\mathrm{red},S}(y\mid v,s,u,\bz).
\label{eq:mean-1a}
\end{equation}

\begin{proposition}
\label{prop:case1a}
Under Assumptions~\ref{ass:L-invariance}--\ref{ass:common},
\begin{equation}
\E(Y_k^{\bar x}\mid U=u,\bZ=\bz)
=m^{\mathrm{red},S}\!\left(
v_k(\bar x),S_k,u,\bz\right).
\label{eq:id-1a}
\end{equation}
\end{proposition}

\begin{proof}
By iterated probability,
\begin{align}
&\mathbb{P}(Y_k^{\bar x}=y\mid U=u,\bZ=\bz)\nonumber\\
&\quad=
\int
\mathbb{P}(Y_k^{\bar x}=y\mid
\Lambda_k^{\bar x}=\lambda,U=u,\bZ=\bz)\,
dF_{\Lambda_k^{\bar x}\mid U,\bZ}(\lambda\mid u,\bz).
\label{eq:proof1a-step1}
\end{align}
Assumption~\ref{ass:L-invariance} gives
\[
F_{\Lambda_k^{\bar x}\mid U,\bZ}
=F_{\Lambda_k\mid U,\bZ},
\]
and Assumption~\ref{ass:full-1a} gives
\[
\mathbb{P}(Y_k^{\bar x}=y\mid
\Lambda_k^{\bar x}=\lambda,U=u,\bZ=\bz)
=q^{\mathrm{full}}\!\left(
y\mid v_k(\bar x),\lambda,u,\bz\right).
\]
Substitution in \eqref{eq:proof1a-step1}, followed first by the
definition \eqref{eq:reduced-1a} and then by
Assumption~\ref{ass:reduced-1a}, yields
\begin{align*}
\mathbb{P}(Y_k^{\bar x}=y\mid U=u,\bZ=\bz)
&=q_k^{\mathrm{red}}\!\left(
y\mid v_k(\bar x),u,\bz\right)\\
&=q^{\mathrm{red},S}\!\left(
y\mid v_k(\bar x),S_k,u,\bz\right).
\end{align*}
Assumption~\ref{ass:common} relates the right-hand side to the observed
schedule at supported values, or supplies the declared extrapolation.
Integrating $y$ gives \eqref{eq:id-1a}.
\end{proof}

\subsection*{Latent variable modeling approach}
\label{sec:state-unaffected}

Instead of using $S_k$ to represent the reduced law after $L$ has been
integrated out, one may specify a model for $L$ itself. A complete
model must first specify the ordering of variables within a time point. Assume the order $L_m,A_m,Y_m$, with $A_m$ fixed to $x_m$ under
intervention $\bar x$. Define
\[
\mathcal G_{m-1}^{L,\bar x}
=\left(\bar L_{m-1}^{\bar x},
\bar Y_{m-1}^{\bar x},\bar x_{m-1}\right),
\qquad m\geq2,
\]
and
\[
\mathcal G_m^{Y,\bar x}
=\left(\bar L_m^{\bar x},
\bar Y_{m-1}^{\bar x},\bar x_m\right),
\qquad m\geq1.
\]
Thus $\mathcal G_{m-1}^{L,\bar x}$ contains all state and outcome values
already generated before $L_m^{\bar x}$, together with the treatment
values fixed through $m-1$. The history
$\mathcal G_m^{Y,\bar x}$ additionally contains $L_m^{\bar x}$ and the
current fixed treatment $x_m$. 

Next, state the initial distribution
\begin{equation}
\mathbb{P}(L_1^{\bar x}=l_1\mid S_1,U=u,\bZ=\bz)
=p_1^L(l_1\mid S_1,u,\bz),
\label{eq:L-initial}
\end{equation}
the later transitions
\begin{equation}
\mathbb{P}(L_m^{\bar x}=l_m\mid
\mathcal G_{m-1}^{L,\bar x},S_m,U=u,\bZ=\bz)
=p^L(l_m\mid L_{m-1}^{\bar x},S_m,u,\bz),
\quad m\geq2,
\label{eq:L-transition-unaffected}
\end{equation}
and the outcome law
\begin{equation}
\mathbb{P}(Y_m^{\bar x}=y_m\mid
\mathcal G_m^{Y,\bar x},S_m,U=u,\bZ=\bz)
=p^Y\!\left(y_m\mid
v_m(\bar x),L_m^{\bar x},S_m,u,\bz\right).
\label{eq:L-outcome-unaffected}
\end{equation}
Equations \eqref{eq:L-transition-unaffected} and
\eqref{eq:L-outcome-unaffected} assert that the shorter arguments on
their right-hand sides are sufficient relative to the explicitly
defined complete histories on their left-hand sides. The same kernels
must govern the observational distribution.

For discrete variables, apply the chain rule using the within-occasion
causal order specified above. Because each $A_m$ is fixed to $x_m$, the
treatment nodes do not contribute probability factors. The result is
\begin{align*}
&\Pr(\bar L_k^{\bar x}=\bar l_k,
\bar Y_k^{\bar x}=\bar y_k\mid U=u,\bZ=\bz)\\
&\quad=
\Pr(L_1^{\bar x}=l_1\mid S_1,U=u,\bZ=\bz)\\
&\qquad{}
\prod_{m=2}^k
\Pr\!\left(L_m^{\bar x}=l_m\mid
\mathcal G_{m-1}^{L,\bar x}
=\left(\bar l_{m-1},\bar y_{m-1},\bar x_{m-1}\right),
S_m,U=u,\bZ=\bz\right)\\
&\qquad{}
\prod_{m=1}^k
\Pr\!\left(Y_m^{\bar x}=y_m\mid
\mathcal G_m^{Y,\bar x}
=\left(\bar l_m,\bar y_{m-1},\bar x_m\right),
S_m,U=u,\bZ=\bz\right).
\end{align*}
Substituting \eqref{eq:L-initial} into the first factor,
\eqref{eq:L-transition-unaffected} into the second group of factors, and
\eqref{eq:L-outcome-unaffected} into the third gives
\begin{align*}
&\Pr(\bar L_k^{\bar x}=\bar l_k,
\bar Y_k^{\bar x}=\bar y_k\mid U=u,\bZ=\bz)\\
&\quad=
p_1^L(l_1\mid S_1,u,\bz)
\prod_{m=2}^k p^L(l_m\mid l_{m-1},S_m,u,\bz)
\prod_{m=1}^k
p^Y\!\left(y_m\mid
v_m(\bar x),l_m,S_m,u,\bz\right).
\end{align*}
Finally, multiply this joint probability by $y_k$ and sum over every
possible counterfactual path. Thus
\begin{align}
&\E(Y_k^{\bar x}\mid U=u,\bZ=\bz)\nonumber\\
&\quad=\sum_{\bar l_k,\bar y_k}y_k\,
p_1^L(l_1\mid S_1,u,\bz)
\prod_{m=2}^k p^L(l_m\mid l_{m-1},S_m,u,\bz)
\prod_{m=1}^k
p^Y\!\left(y_m\mid v_m(\bar x),l_m,S_m,u,\bz\right).
\label{eq:L-gformula-unaffected}
\end{align}
Sums would be replaced by integrals for continuous variables.

The factors involving $Y_1,\ldots,Y_{k-1}$ are retained in
\eqref{eq:L-gformula-unaffected} to display the complete sequential
factorization. Under the particular transition models above, earlier
outcomes do not enter the latent variable or outcome kernels. If
\[
m^Y(v,l,s,u,\bz)
=\sum_y y\,p^Y(y\mid v,l,s,u,\bz),
\]
then the same formula can therefore be written more compactly as
\[
\E(Y_k^{\bar x}\mid U=u,\bZ=\bz)
=\sum_{\bar l_k}
m^Y\!\left(v_k(\bar x),l_k,S_k,u,\bz\right)
p_1^L(l_1\mid S_1,u,\bz)
\prod_{m=2}^k p^L(l_m\mid l_{m-1},S_m,u,\bz).
\]

\end{document}